\documentclass[lettersize, journal]{IEEEtran}

\usepackage{amsmath,amsfonts}
\usepackage{algorithmic}
\usepackage{algorithm}
\usepackage{array}
\usepackage[caption=false,font=small,labelfont=sf,textfont=sf]{subfig}
\usepackage{textcomp}
\usepackage{stfloats}
\usepackage{url}
\usepackage{verbatim}
\usepackage{graphicx}
\usepackage{cite}
\usepackage[colorlinks, linkcolor=blue, urlcolor=black, citecolor=green]{hyperref}
\usepackage{color}
\usepackage{makecell}
\usepackage{booktabs}
\usepackage{soul}
\usepackage{multirow}
\usepackage{tcolorbox}
\usepackage{wrapfig}
\usepackage{tcolorbox}

\allowdisplaybreaks[4]

\newtheorem{theorem}{\bf Theorem}

\begin{document}

\title{A Joint Spectro-Temporal Relational Thinking Based Acoustic Modeling Framework}

\author{
	Zheng Nan, 
	Ting Dang,~\IEEEmembership{Member,~IEEE},
	Vidhyasaharan Sethu,~\IEEEmembership{Member,~IEEE}, and
	Beena Ahmed,~\IEEEmembership{Member,~IEEE}
\thanks{
	Z. Nan, V. Sethu, and B. Ahmed are with the School of Electrical Engineering and Telecommunications, University of New South Wales, Sydney, NSW 2052, Australia (e-mail: \{zheng.nan, v.sethu, beena.ahmed\}@unsw.edu.au).
	}
\thanks{
    T. Dang is with the School of Computing and Information Systems, University of Melbourne, and is also affiliated with the School of Electrical Engineering and Telecommunications, University of New South Wales, Sydney, NSW 2052, Australia (e-mail: ting.dang@unimelb.edu.au).
	}
}



\maketitle

\begin{abstract}
Relational thinking refers to the inherent ability of humans to form mental impressions about relations between sensory signals and prior knowledge, and subsequently incorporate them into their model of their world. Despite the crucial role relational thinking plays in human understanding of speech, it has yet to be leveraged in any artificial speech recognition systems. Recently, there have been some attempts to correct this oversight, but these have been limited to coarse utterance-level models that operate exclusively in the time domain. In an attempt to narrow the gap between artificial systems and human abilities, this paper presents a novel spectro-temporal relational thinking based acoustic modeling framework. Specifically, it first generates numerous probabilistic graphs to model the relationships among speech segments across both time and frequency domains. The relational information rooted in every pair of nodes within these graphs is then aggregated and embedded into latent representations that can be utilized by downstream tasks.
Models built upon this framework outperform state-of-the-art systems with a 7.82\% improvement in phoneme recognition tasks over the TIMIT dataset. In-depth analyses further reveal that our proposed relational thinking modeling mainly improves the model's ability to recognize vowels, which are the most likely to be confused by phoneme recognizers.
\end{abstract}


\section{Introduction}\label{sec.introduction}
\noindent
Deep learning techniques have brought in substantial advancements into automatic speech recognition (ASR), making it one of the most promising means of human-machine communication \cite{hinton2012deep}. However, most deep neural network (DNN) based speech recognition systems \cite{vinyals2012revisiting, abdel2014convolutional, chan2016listen, passricha2019hybrid, wang2020transformer, baevski2020wav2vec, gulati2020conformer} have drawn limited inspiration from the way speech is processed and recognized by human brain \cite{bohnstingl2022speech}, instead treating the process as a black-box. As a consequence, the performances of these systems still lag behind that of the human brain \cite{malik2021automatic}. Recognizing the limitations inherent in current artificial systems, recently researchers have endeavored to integrate biologically inspired mechanisms into existing DNN based systems, seeking to enhance interpretability and narrow the gap between artificial systems and the human brain \cite{dong2020cif, bohnstingl2022speech}.

\begin{figure}[]
\centering
\includegraphics[width=0.49\textwidth]{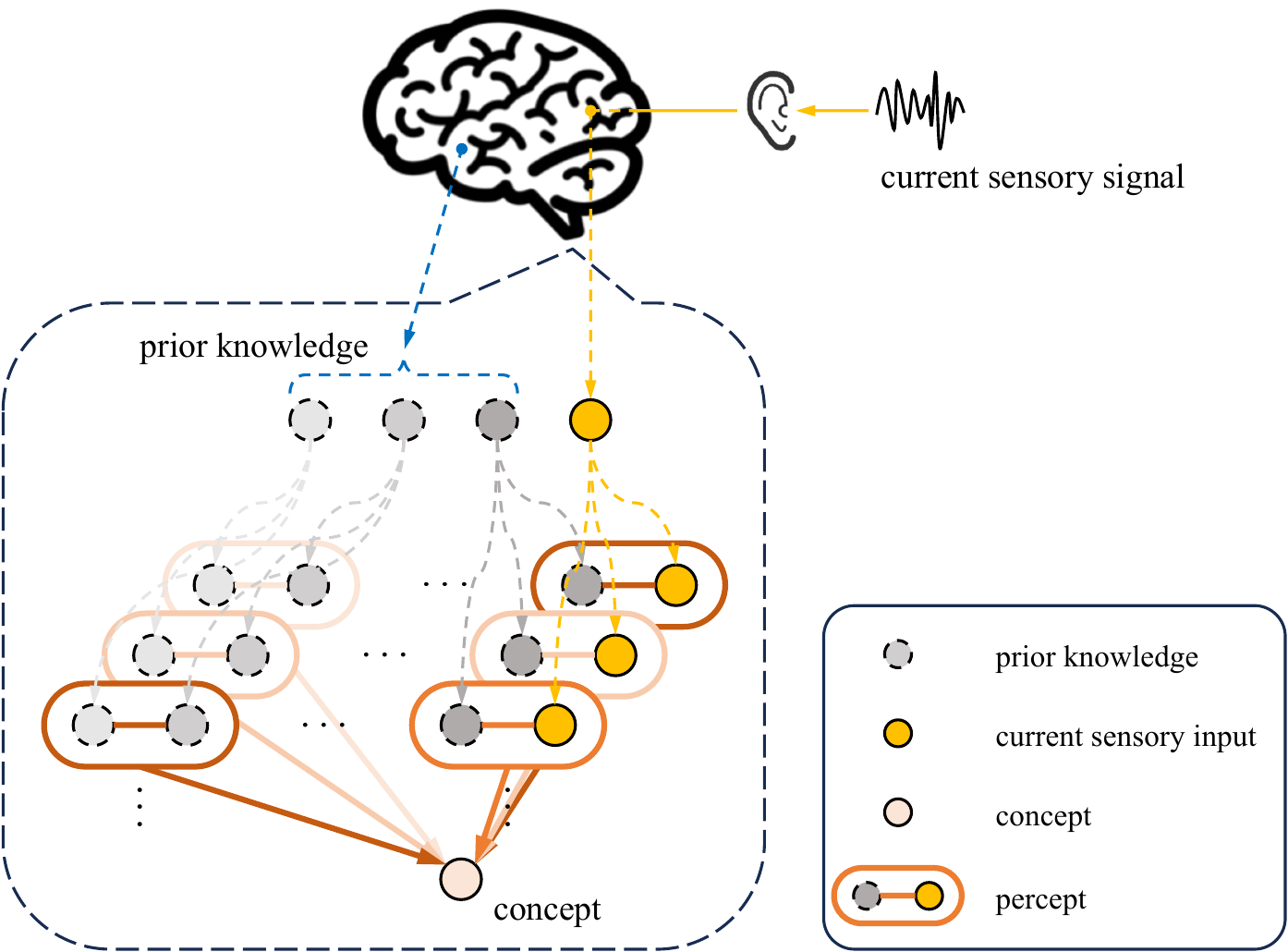}
\captionsetup{font=small}
\caption{An illustration of relational thinking process in the human brain. While listening, one's mind is continuously and unconsciously filled with innumerable percepts pertaining to relations between current sensory signals and prior knowledge. These perceptions are then aggregated and transformed into concepts.}\label{fig.human_relational_thinking}
\end{figure}

\begin{figure}[]
\centering
\includegraphics[width=0.49\textwidth]{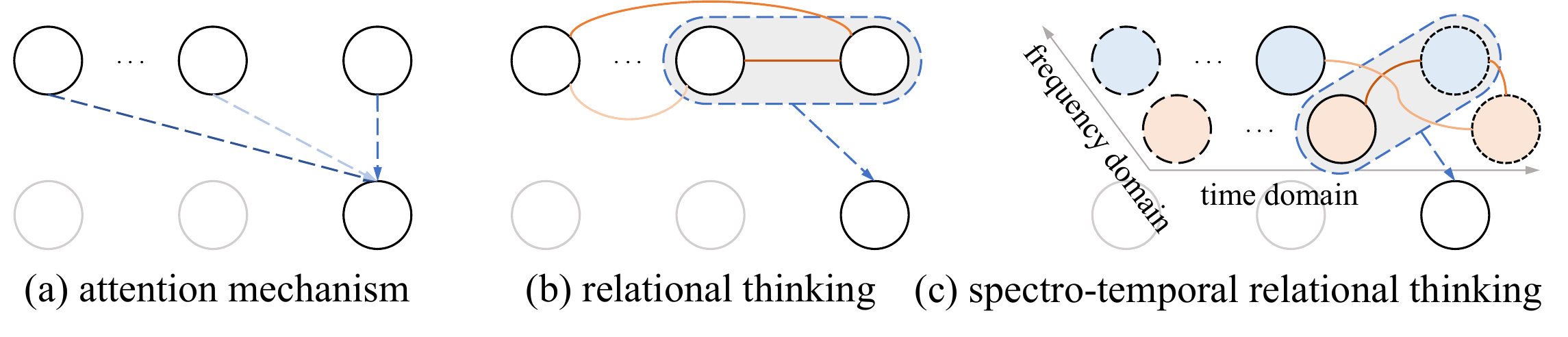}
\captionsetup{font=small}
\caption{Graphical illustrations of (a) attention mechanism, (b) conventional relational thinking, and (c) proposed joint spectro-temporal relational thinking. A darker line indicates greater importance (i.e., with a larger weight).}\label{fig.attention_vs_relational_thinking}
\end{figure}

The human brain employs an inherent {\it relational thinking} process for speech recognition and comprehension \cite{birjandi2012review}. This is a fundamental human learning process that enables discerning meaningful patterns within the continuous stream of sensory signals \cite{alexander2016relational}. Specifically, while one is hearing, seeing, smelling, etc., their mind is continuously and unconsciously filled with innumerable mental impressions that pertain to relations between current sensory signals and prior knowledge \cite{peirce2012philosophical}. These mental impressions, or {\it percepts}, are then by some means aggregated and transformed into generalized understandings, or {\it concepts}; as illustrated in Fig.~\ref{fig.human_relational_thinking}. 
Most state-of-the-art systems, e.g., wav2vec2 \cite{baevski2020wav2vec}, use transformer architectures \cite{vaswani2017attention}, which employ attention mechanisms to capture dependencies between different parts of the sequence. However, these systems do not explicitly model the relational information inherent in the sequence in the same way as the human brain. 
The attention mechanisms instead assess the significance of different parts of the sequential input, 
allowing the model to focus on only pertinent information, 
as illustrated by Fig.~\ref{fig.attention_vs_relational_thinking}~(a). In contrast, relational thinking captures the inherent relationships and interactions between various pair-wise elements or features within the input sequence and estimates each entry of the output by aggregating all the pair-wise information, as shown in Fig.~\ref{fig.attention_vs_relational_thinking}~(b). 
Relational thinking thus better  learns the implications of co-occurring pairs of informative elements. This proves particularly beneficial for speech recognition, as certain pairs tend to appear jointly, for instance, the phonemes /m/ and /iy/ (``\underline{me}'', ``autono\underline{my}'', etc.),  knowledge not intrinsically captured by attention based models.

One of the few examples of the use of relational thinking models was proposed in a conversational speech recognition system \cite{huang2020deep}, 
where the acquired relational information was utilized as an additional input in the recognition task. 
However, \cite{huang2020deep} only investigated utterance-level relational information in conversational scenarios, and this approach is only applicable to cases where the input and output sequences have the same length. In another example, from the domain of natural language processing, \cite{xue2021gdpnet} proposed predicting the relation type of two entities by extracting relationships between words. Both \cite{huang2020deep} and \cite{xue2021gdpnet} modeled the relations either at the utterance-level or the word-level. However, humans also process speech and language at the more granular level of phonemes \cite{dusan2005integrating, wingfield2017relating}. Furthermore, existing works have modeled the relations among elements of the input sequence separated in time only, whereas humans process speech by jointly considering multiple domains (e.g., time, frequency, semantics, etc.) rather than focusing exclusively on relationships in the time domain \cite{10.5555/555733}. 

In this paper, i)~we first identify the limitations of self-attention mechanism in mimicking human brain by comparing it with relational thinking, where we highlight the inherent differences between the types of information captured by these two processes. ii)~Then, we propose a novel joint spectro-temporal relational thinking based acoustic modeling framework. 
This framework captures relationships across both time and frequency domains of the sensory input (as illustrated by Fig.~\ref{fig.attention_vs_relational_thinking}~(c)), in contrast to previous approaches that focus solely on temporal patterns.
iii)~A tractable loss that optimizes the variational lower bound for the model log-likelihood is developed to tackle real-world scenarios where the input and output sequences differ in length.
iv)~Models built upon our proposed framework outperform the state-of-the-art baseline. 
Further analysis shows that the performance gain primarily originates from the model's enhanced ability to recognize vowels. This enhancement mirrors human proficiency in recognizing vowels more readily than consonants \cite{meyer2006human}. We also investigate the relevance of the captured relations to phoneme groups, where the patterns involved in the relations exhibit more similarities for phoneme classes within the same group. 
Additionally, the generalizability of the proposed framework is validated by employing other types of acoustic features (e.g., MFCCs), where relational thinking modeling consistently benefits downstream tasks.



\section{Modeling Relational Thinking}\label{sec.relational_thinking}
\begin{figure*}[]
\centering
\includegraphics[width=\textwidth]{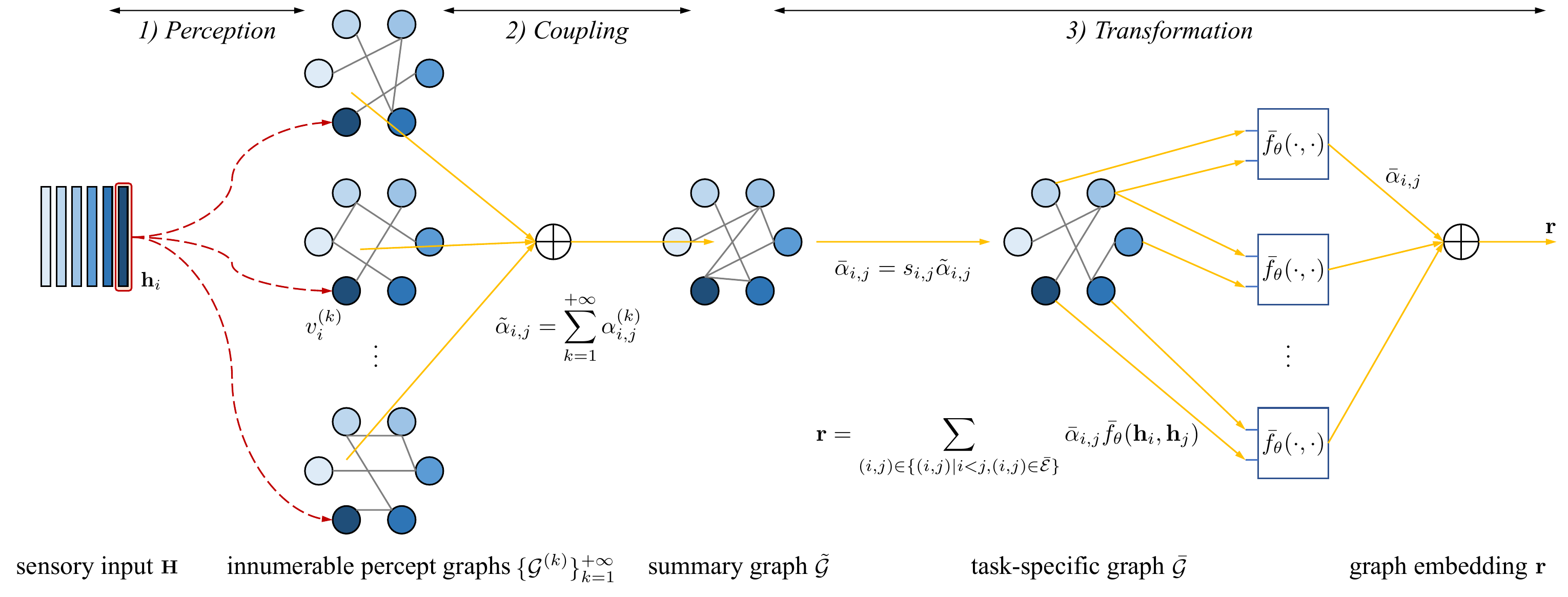}
\captionsetup{font=small}
\caption{Modeling of relational thinking process.}\label{fig.relational_thinking}
\end{figure*}

Previous relational thinking approaches have employed graphs to model relationships between entries (or time steps) of a sequence, where each entry has been regarded as a node in the graphs. The goal of such approaches is to capture meaningful pair-wise patterns over time using these graphs (as illustrated by Fig.\ref{fig.attention_vs_relational_thinking}~(b)), and then aggregate and transform the relational information involved in the graphs into a latent form that can be interpreted by subsequent layers of the model.

Consider a sequence of acoustic features $\mathbf{H} = \left [ \mathbf{h}_1, \ldots, \mathbf{h}_T \right ]$ 
corresponding to $T$ time steps. As illustrated by Fig.~\ref{fig.relational_thinking}, the relational thinking process is carried out via the following three steps \cite{huang2020deep}:

\subsubsection{Perception}
We first construct an infinite number of graphs $\{ \mathcal{G}^{(k)} \}_{k = 1}^{+\infty}$, where $\mathcal{G}^{(k)} ( \mathcal{V}^{(k)}, \mathcal{E}^{(k)} )$ is the $k$-th {\it percept} graph, with $\mathcal{V}^{(k)}$ and $\mathcal{E}^{(k)}$ denoting the node set and edge set, respectively. Each $\mathbf{h}_i \in \mathbb{R}^{D_h}, i = 1, \ldots, T$ corresponds to a node $v_i^{(k)}$ in every percept graph $\mathcal{G}^{(k)}$, while each element $\alpha_{i, j}^{(k)}$ of the adjacency matrix $\mathbf{A}^{(k)}$ is associated with an edge $e_{i, j}^{(k)} \in \mathcal{E}^{(k)}$ between a pair of nodes $(v_i^{(k)}, v_j^{(k)})$ of $\mathcal{G}^{(k)}$. The value of $\alpha_{i, j}^{(k)}$ indicates the significance of the co-occurrence of node pair $(v_i^{(k)}, v_j^{(k)})$.

Since the percepts form at an unconscious level of awareness \cite{rapp2023processing}, we assume that the probability of an edge's existence within the percept graphs is close to zero.
To model this characteristic, we let the edge weights for the percept graphs follow a set of Bernoulli distributions, i.e., 
\begin{equation}\label{eq.bern}
\left \{ \alpha_{i, j}^{(k)} \right \}_{k = 1}^{+\infty} \sim \text{Bern} (\lambda_{i, j}),
\end{equation}
where the probability of edge existence $\lambda_{i, j} \rightarrow 0$.


\subsubsection{Coupling}
Coupling aims to obtain an equivalent {\it summary} graph $\tilde{\mathcal{G}}$, which is capable of representing the infinite number of percept graphs $\{ \mathcal{G}^{(k)} \}_{k = 1}^{+\infty}$. In this graph, the original nodes $\mathbf{h}_1, \ldots, \mathbf{h}_T$ are preserved. While since it is intractable to simply take a summation over all adjacency matrices $\{ \mathbf{A}^{(k)} \}_{k = 1}^{+\infty}$, each edge $\tilde{\alpha}_{i, j}$ of $\tilde{\mathcal{G}}$ is equivalently generated upon sampling from a Binomial distribution, i.e.,
\begin{equation}\label{eq.binomial}
\tilde{\alpha}_{i, j} \sim \mathcal{B}(n, \lambda_{i, j}) \quad \Leftrightarrow \quad \tilde{\alpha}_{i, j} = \sum_{k = 1}^{+\infty} \alpha_{i, j}^{(k)},
\end{equation}
where $n \rightarrow +\infty$ and $\lambda_{i, j} \rightarrow 0$. However, due to the intractability of $n$ and $\lambda_{i, j}$, we cannot directly draw samples $\tilde{\alpha}_{i, j}$ from $\mathcal{B}(n, \lambda_{i, j})$. To tackle this intractability, we adopt the following theorem \cite{huang2020deep}:
\begin{theorem}\label{theorem_1}
{\it Let $\mathcal{B}(n, \lambda)$ denote a Binomial distribution with $n \rightarrow +\infty, \lambda \rightarrow 0$, and let $m = n \lambda$. There exists a Gaussian distribution $\mathcal{N}(m, m(1 - m))$ that approximates $\mathcal{B}(n, \lambda)$ with a bounded approximation error, where
\begin{equation}\label{eq.m}
m = \frac{1}{2} \left \{ 1 + \frac{ 2 \sigma^2 }{ 1 - 2 \mu } - \left [ 1 + \left ( \frac{ 2 \sigma^2 }{ 1 - 2 \mu } \right )^2 \right ]^{{1} / {2}} \right \}
\end{equation}
is derived from a Gaussian distribution $\mathcal{N} \left ( \mu, \sigma^2 \right )$ with $\mu < 1 / 2$.
}
\end{theorem}

According to Theorem~\ref{theorem_1}, by letting $m_{i, j} = n \lambda_{i, j}$, we can bypass the direct parameterization of both the infinite $n$ and the near-zero $\lambda_{i, j}$, and find a tractable Gaussian proxy $\mathcal{N}(m_{i, j}, m_{i, j}(1 - m_{i, j}))$ for the original Binomial distribution $\mathcal{B}(n, \lambda_{i, j})$ in (\ref{eq.binomial}), from which we can draw samples $\tilde{\alpha}_{i, j}$.

\subsubsection{Transformation}
Transformation converts the innumerable unconscious percepts into a recognizable notion of knowledge. 
Therefore, the summary graph $\tilde{\mathcal{G}}$, which represents the infinite number of percept graphs, is first transformed into a {\it task-specific} graph $\bar{\mathcal{G}}$, and an informative representation $\mathbf{r}$ is subsequently abstracted from $\bar{\mathcal{G}}$ for downstream tasks. 

Specifically, this transformation is designed as first weighting each edge $\tilde{\alpha}_{i, j}$ of $\tilde{\mathcal{G}}$ with a Gaussian variable $s_{i, j}$:
\begin{equation}\label{eq.bar_alpha}
\bar{\mathbf{A}} = \mathbf{S} \odot \tilde{\mathbf{A}},
\end{equation}
where $\odot$ denotes the Hadamard product, $\mathbf{S}$ is the graph transformation matrix collecting $s_{i, j}$, $\tilde{\mathbf{A}}$ and $\bar{\mathbf{A}}$ are the adjacency matrices of $\tilde{\mathcal{G}}$ and $\bar{\mathcal{G}}$, respectively. $s_{i, j}$ is assumed to be conditioned on the corresponding edge $\tilde{\alpha}_{i, j}$ of $\tilde{\mathcal{G}}$, i.e.,
\begin{equation}\label{eq.s}
s_{i, j} | \tilde{\alpha}_{i, j} \sim \mathcal{N} \left ( \tilde{\alpha}_{i, j} \mu_{i, j},  \tilde{\alpha}_{i, j} \sigma_{i, j}^2 \right ).
\end{equation}
Next, a graph embedding $\mathbf{r}$ is abstracted from $\bar{\mathcal{G}}$ by summing up the embeddings of all node pairs weighted by $\bar{\alpha}_{i, j}$ as:
\begin{equation}\label{eq.graph_embedding}
\mathbf{r} = \sum_{(i, j) \in \left \{ (i, j) | i < j, (i, j) \in \bar{\mathcal{E}} \right \}} \bar{\alpha}_{i, j} \bar{f}_{\theta}(\mathbf{h}_{i}, \mathbf{h}_{j}),
\end{equation}
where $\bar{\mathcal{E}}$ is the edge set of $\bar{\mathcal{G}}$, and $\bar{f}_{\theta}(\cdot, \cdot)$ denotes a node pair embedding function \cite{kipf2018neural}. As indicated by \eqref{eq.graph_embedding}, $\mathbf{r}$ captures the importance of the co-occurrence of entry pairs within the input $\mathbf{H}$. 
This knowledge is to be used as additional input for downstream tasks.

\section{Self-attention vs. Relational Thinking}\label{app.vs_self_attention}
To form a deeper view of the unique information captured by relational thinking, in addition to what is already provided by self-attention mechanism \cite{vaswani2017attention}, we compared these two processes.

In the self-attention mechanism, the weights and the representation of the relations involved are calculated as
\begin{subequations}\label{eq.attention}
\begin{align}
\nonumber \alpha_{i, j} = & \text{softmax}( \text{score}( \mathbf{W}_q \mathbf{h}_i, \mathbf{W}_k \mathbf{h}_j ) ), \\
\label{eq.attention_a} & ~~~~~~~~~~~~~~~~~~~~~~~~~~~ i, j = t - w + 1, \ldots, t, \\
\label{eq.attention_b} \mathbf{e}_{i} = & \sum_{j = t - w + 1}^{t} \alpha_{i, j} f_v(\mathbf{h}_{j}), \quad i = t - w + 1, \ldots, t, \\
\label{eq.attention_c} \mathbf{r}_{t} = & \mathbf{e}_{t},
\end{align}
\end{subequations}
where $\text{score}(\mathbf{q}, \mathbf{k}) = \mathbf{k}^T \mathbf{q} / \sqrt{|\mathbf{k}|}$, and $f_v(\mathbf{h}) = \mathbf{W}_v \mathbf{h}$ \cite{vaswani2017attention}. For the ease of comparison, we reframe the relational thinking process (\ref{eq.bern})--(\ref{eq.graph_embedding}) into the following form:
\begin{subequations}\label{eq.relational_thinking}
\begin{align}
\label{eq.relational_thinking_b} \mathbf{e}_{i}^{(t)} = & \sum_{j = t - w + 1, j > i}^{t} \bar{\alpha}_{i, j}^{(t)} \bar{f}_{\theta}(\mathbf{h}_{i}, \mathbf{h}_{j}), \quad i = t - w + 1, \ldots, t, \\
\label{eq.relational_thinking_c} \mathbf{r}_{t} = & \sum_{i = t - w + 1}^t \mathbf{e}_{i}^{(t)},
\end{align}
\end{subequations}
where $\bar{\alpha}_{i, j}^{(t)}$ is obtained by a generative process (\ref{eq.bern})--(\ref{eq.s}), and $\bar{f}_{\theta}(\mathbf{h}_i, \mathbf{h}_j) = \text{MLP} ( [ \mathbf{h}_i^T, \mathbf{h}_j^T ]^T )$.

As indicated by (\ref{eq.attention}) and (\ref{eq.relational_thinking}), both the self-attention mechanism and relational thinking share a similar calculation structure. With either of these techniques, we can eventually derive a representation $\mathbf{r}_t$ that captures the relations involved in different nodes. This representation $\mathbf{r}_t$ has a form of a weighted sum of embeddings.

However, the methods for determining the weights $\alpha_{i, j}$ in (\ref{eq.attention_a}) and $\bar{\alpha}_{i, j}^{(t)}$ differ between the two techniques. In self-attention mechanism, a node $\mathbf{h}_i$ is initially projected into a query space and a key space by $\mathbf{W}_q$ and $\mathbf{W}_k$, respectively. Then, there is a scoring process $\text{score}(\cdot, \cdot)$, where the score typically quantifies the relevance or importance of the key vector $\mathbf{W}_k \mathbf{h}_j$ concerning the query vector $\mathbf{W}_q \mathbf{h}_i$. While in relational thinking, the weights $\bar{\alpha}_{i, j}^{(t)}$ are generated through a generative process (\ref{eq.bern})--(\ref{eq.s}). Each weight corresponds to an edge connecting two nodes in the task-specific graph.

Next, when comparing (\ref{eq.attention_b}) and (\ref{eq.relational_thinking_b}), we can further observe notable differences in the embedding functions used by the two techniques. In self-attention mechanism, a single node is typically embedded using a linear transformation $f_v(\mathbf{h}) = \mathbf{W}_v \mathbf{h}$. In contrast, in relational thinking, a pair of nodes is embedded together using a network $\bar{f}_{\theta}(\cdot, \cdot)$. This distinction leads to a fundamental difference in the outcomes. Specifically, self-attention mechanism ultimately calculates a weighted sum of node embeddings, while relational thinking computes a weighted sum of node pair embeddings. 

We further show that even the stacked self-attention mechanism (i.e., multiple layers of self-attention) and relational thinking are distinct processes.
Without loss of generality, consider a simplified 2-layer self-attention network, where each layer $l$ comprises two nodes $\mathbf{h}_1^{(l)}$ and $\mathbf{h}_2^{(l)}$. According to (\ref{eq.attention}), the calculation of nodes in the subsequent layer $l + 1$ is as follows:
\begin{equation}
\left [ \mathbf{h}_1^{(l + 1)}, \mathbf{h}_2^{(l + 1)} \right ] = \mathbf{W}_v^{(l)} \left [ \mathbf{h}_1^{(l)}, \mathbf{h}_2^{(l)} \right ] \left [ 
\begin{array}{cc}
\alpha_{1, 1}^{(l)} & \alpha_{2, 1}^{(l)} \\
\alpha_{1, 2}^{(l)} & \alpha_{2, 2}^{(l)} 
\end{array}
\right ].
\end{equation}
As a result, the state of a node $\mathbf{h}_2^{(3)}$ after undergoing two layers of self-attention calculations is 
\begin{subequations}
\begin{align}
\nonumber \mathbf{h}_2^{(3)} = & \alpha_{2, 1}^{(2)} \mathbf{W}_v^{(2)} \left ( \alpha_{1, 1}^{(1)} \mathbf{W}_v^{(1)} \mathbf{h}_1^{(1)} + \alpha_{1, 2}^{(1)} \mathbf{W}_v^{(1)} \mathbf{h}_2^{(1)} \right ) + \\
\label{eq.stacked_att} & \alpha_{2, 2}^{(2)} \mathbf{W}_v^{(2)} \left ( \alpha_{2, 1}^{(1)} \mathbf{W}_v^{(1)} \mathbf{h}_1^{(1)} + \alpha_{2, 2}^{(1)} \mathbf{W}_v^{(1)} \mathbf{h}_2^{(1)} \right ) \\
\nonumber = & \left ( \alpha_{1, 1}^{(1)} \alpha_{2, 1}^{(2)} + \alpha_{2, 1}^{(1)} \alpha_{2, 2}^{(2)} \right ) \mathbf{W}_v^{(2)} \mathbf{W}_v^{(1)} \mathbf{h}_1^{(1)} + \\
\label{eq.stacked_att_2} & \left ( \alpha_{1, 2}^{(1)} \alpha_{2, 1}^{(2)} + \alpha_{2, 2}^{(1)} \alpha_{2, 2}^{(2)} \right ) \mathbf{W}_v^{(2)} \mathbf{W}_v^{(1)} \mathbf{h}_2^{(1)}.
\end{align}
\end{subequations}
Even though (\ref{eq.stacked_att}) may exhibit a similar form to (\ref{eq.graph_embedding}), particularly when we view $\mathbf{W}_v^{(2)} ( \alpha_{1, 1}^{(1)} \mathbf{W}_v^{(1)} \mathbf{h}_1^{(1)} + \alpha_{1, 2}^{(1)} \mathbf{W}_v^{(1)} \mathbf{h}_2^{(1)} )$ and $\mathbf{W}_v^{(2)} ( \alpha_{2, 1}^{(1)} \mathbf{W}_v^{(1)} \mathbf{h}_1^{(1)} + \alpha_{2, 2}^{(1)} \mathbf{W}_v^{(1)} \mathbf{h}_2^{(1)} )$ as node pair embedding functions from a linear family $\mathcal{F} ( \mathbf{h}_1^{(1)}, \mathbf{h}_2^{(1)} )$, it is crucial to note that $\mathbf{h}_2^{(3)}$ is fundamentally still a weighted sum of node embeddings (as revealed by (\ref{eq.stacked_att_2})) rather than a weighted sum of node pair embeddings as obtained by relational thinking (\ref{eq.graph_embedding}), where $\bar{f}_{\theta}(\mathbf{h}_{i}, \mathbf{h}_{j})$ is an arbitrary node pair embedding function. Therefore, unlike relational thinking, the stacked self-attention mechanism cannot effectively assess the importance of a pair of nodes that covary. 

This all shows that relational thinking provides additional information about speech not available from the self-attention mechanism. As described in Section~\ref{sec.relational_thinking}, existing works have modeled the relational thinking process in only the time domain \cite{huang2020deep, xue2021gdpnet}. However, speech cannot be sufficiently characterized using time domain information alone, as the information involved in other domains (e.g., frequency domain) is also crucial to the ultimate task. Therefore, we propose modeling the relational thinking process jointly across multiple domains. This aligns with how human brain processes speech, and will enable a more comprehensive description of speech signals \cite{10.5555/555733}.

\section{Proposed Spectro-temporal Relational Thinking Framework}\label{sec.proposed}

To exploit the range of information that are more readily accessible from different domains, (e.g., time domain, frequency domain, etc.),
in this section we propose an acoustic modeling framework that models the relational thinking process jointly across both the time and frequency domains (and more generally across the dimensions of any acoustic representation). 

\subsection{Spectro-temporal Relational Thinking Based Acoustic Modeling}\label{seq.wav2vec2_with_rt}
\begin{figure}[]
\centering
\includegraphics[width=0.48\textwidth]{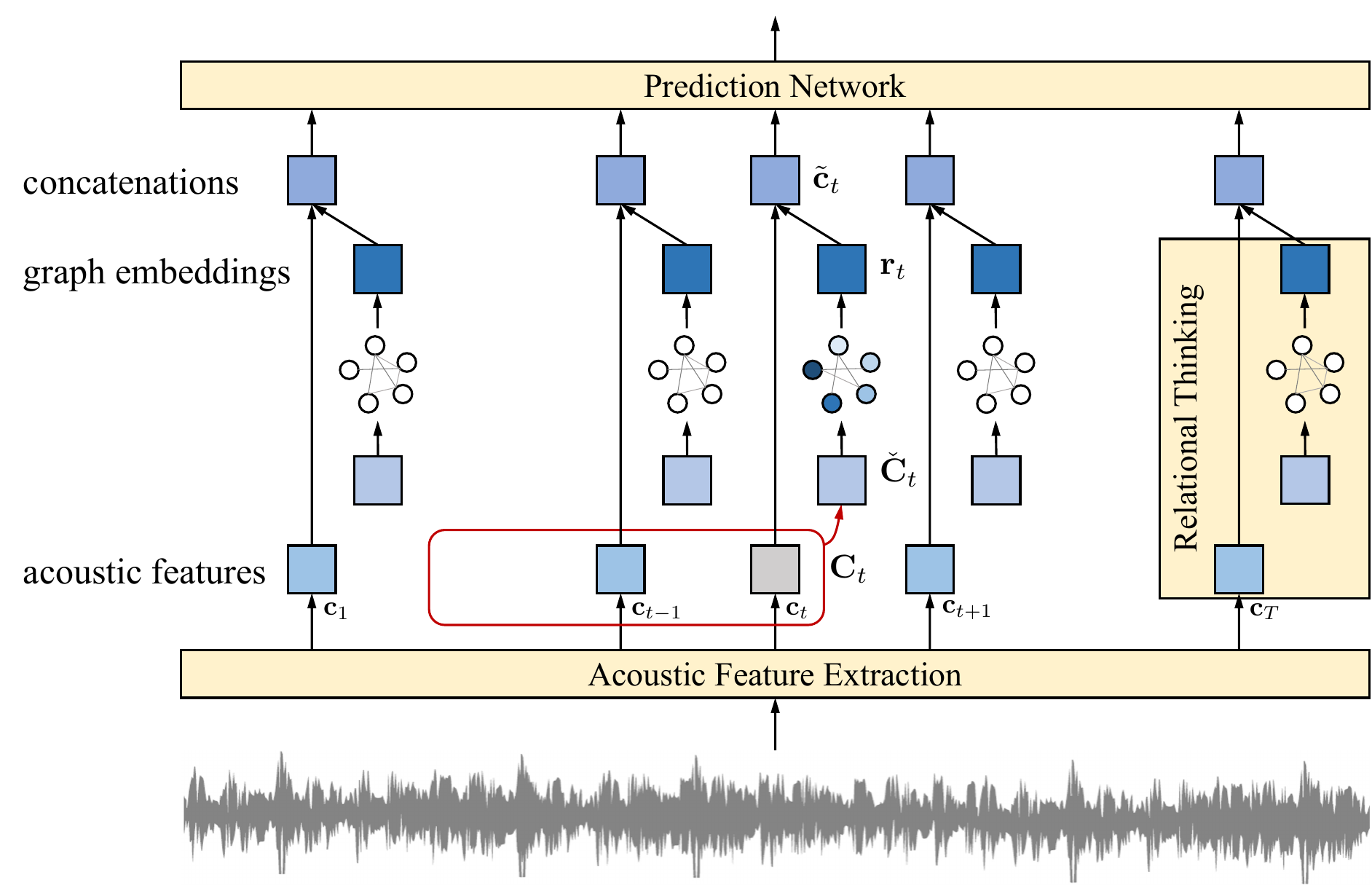}
\captionsetup{font=small}
\caption{Spectro-temporal relational thinking based acoustic modeling framework.}\label{fig.wav2vec2_with_relational_thinking}
\end{figure}

The structure of the proposed acoustic modeling framework is depicted in Fig.~\ref{fig.wav2vec2_with_relational_thinking}. Given the raw waveform of a speech, we first employ the feature extraction module to calculate the acoustic feature vectors $\mathbf{c}_t \in \mathbb{R}^{D_c}, t = 1, \ldots, T$ corresponding to each of the time steps. Then, we re-organize them into a set of feature maps $\mathcal{C} = \{ \mathbf{C}_1, \ldots, \mathbf{C}_T \}$ by forming each feature map with the current and the previous $w - 1$ time steps as $\mathbf{C}_t = [\mathbf{c}_{t - w + 1}, \ldots, \mathbf{c}_t]$\footnote{In a slight abuse of terminology, we refer to the feature space, in which $\mathbf{c}_1, \ldots, \mathbf{c}_T$ exist, as a {\it frequency} domain, although $\mathbf{c}_t$ can be an arbitrary type of acoustic feature.}, guaranteeing the incorporation of causality. $\mathbf{C}_t$ is subsequently used 
as the sensory input for relational thinking modeling of time step $t$. 
For time steps with $t < w$, specifically, $\mathbf{C}_t$ is padded with $\mathbf{0} \in \mathbb{R}^{D_c}$ such that all feature maps $\mathbf{C}_t, \forall t$ have the identical dimension of $D_c \times w$. 

\begin{figure*}[]
\centering
\includegraphics[width=0.95\textwidth]{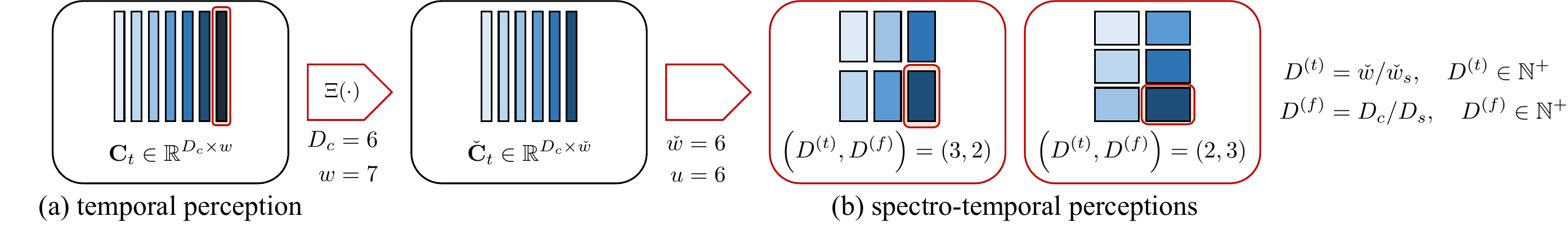}
\captionsetup{font=small}
\caption{Perception step of (a) temporal and (b) spectro-temporal relational thinking modeling.}\label{fig.time_vs_time_and_feature}
\end{figure*}

For the relational thinking module, every $\mathbf{C}_t$ is first smoothed and sub-sampled as
\begin{equation}\label{eq.sub_sample}
\check{\mathbf{C}}_t = \Xi(\mathbf{C}_t),
\end{equation}
where $\Xi$ denotes a filtering operator. The function of $\Xi$ is to adjust the dimension of the original feature map $\mathbf{C}_t$, such that the resultant $\check{\mathbf{C}}_t$ has a dimension suitable for the subsequent spectro-temporal relational thinking modeling. Next, $\check{\mathbf{C}}_t$ is divided into a number of sub-feature maps as  
\begin{equation}
\check{\mathbf{C}}_t = \left [ 
\begin{array}{c c c}
\mathbf{\Lambda}_{t, 1, 1} & \cdots & \mathbf{\Lambda}_{t, 1, D^{(t)}} \\
\vdots & \ddots & \vdots \\
\mathbf{\Lambda}_{t, D^{(f)}, 1} & \cdots & \mathbf{\Lambda}_{t, D^{(f)}, D^{(t)}} \\
\end{array}
\right ],
\end{equation}
where $\check{\mathbf{C}}_t \in \mathbb{R}^{D_c \times \check{w}}$. Every one of the total $u = D^{(f)} \times D^{(t)}$ sub feature maps $\mathbf{\Lambda}_{t, d^{(f)}, d^{(t)}} \in \mathbb{R}^{D_s \times \check{w}_s}, d^{(f)} = 1, \ldots, D^{(f)}, d^{(t)} = 1, \ldots, D^{(t)}$ spans across both time and frequency domains, and is ready to be mapped to a node within the percept graphs $\mathcal{G}^{(k)}_t$.
As for the filtering $\Xi$ in (\ref{eq.sub_sample}), we explain its necessity with the example in Fig.~\ref{fig.time_vs_time_and_feature}. For the perception step of time domain modeling illustrated by Fig.~\ref{fig.time_vs_time_and_feature}~(a), each $\mathbf{c}_t$ from a time step can be directly mapped to a node in the percept graphs, with the number of nodes in a graph corresponding to the number of time steps ($w=7$) included in $\mathbf{C}_t$. However, as per the spectro-temporal modeling, each node in the percept graphs encompasses information in both the time and frequency domains, spanning over $\check{w}_s$ and $D_s$, respectively. As illustrated by Fig.~\ref{fig.time_vs_time_and_feature}~(b), given $D_c = 6$, $w = 7$, and $u = 6$, it is not possible to evenly divide the $6 \times 7$ feature map $\mathbf{C}_t$ into 2 rows and 3 columns, or 3 rows and 2 columns of sub-feature maps in the two red blocks in the figure. As a result, adjustments for the dimension of the original feature map $\mathbf{C}_t$ is necessary. We implement $\Xi$ with a temporal convolution in the proposed framework.
Furthermore, given $\check{\mathbf{C}}_t$, $D^{(t)}$ and $D^{(f)}$ (as defined in Fig.~\ref{fig.time_vs_time_and_feature}) in fact determine the {\it resolutions} of relational thinking modeling in time and frequency domains, respectively. A higher resolution $D^{(t)}$ or $D^{(f)}$ indicates a more fine-grained capture of relations across the corresponding domain. For a given number of sub feature maps $u$ to be mapped to the nodes within the graphs, there exist multiple choices for the resolutions $(D^{(t)}, D^{(f)})$. As illustrated by the two solutions for the example in Fig.~\ref{fig.time_vs_time_and_feature}~(b), given $u = 6$, we can obtain either $(D^{(t)}, D^{(f)}) = (3, 2)$ or $(D^{(t)}, D^{(f)}) = (2, 3)$ for the spectro-temporal perception. Variations in the resolution settings can have different effects on the performance of the downstream task. This aspect will be discussed in detail in Section~\ref{sec.performance}.

By sequentially performing the perception, coupling, and transformation steps (\ref{eq.bern})--(\ref{eq.graph_embedding}) toward $\check{\mathbf{C}}_t$ for each time step $t$, we can obtain a sequence of graph embeddings $\mathbf{r}_1, \ldots, \mathbf{r}_T$. 
Unlike the time domain modeling discussed in Section~\ref{sec.relational_thinking}, where a node pair refers to the co-occurrence of two temporal frames, here each node pair represents a spectro-temporal {\it pattern} formed by the co-occurrence of two sub feature maps either within an {\it interval} (i.e., the temporal span covered by a sub feature map $\mathbf{\Lambda}_{t, d^{(f)}, d^{(t)}}$) or across intervals. Therefore, by incorporating both time and frequency domains, the graph embeddings $\mathbf{r}_t$ are able to capture not only the relations between time intervals, but also the relations across different {\it frequency bands} within an interval or across intervals.


By concatenating each $\mathbf{r}_t$ with the corresponding acoustic feature vector $\mathbf{c}_t$, we then obtain a more comprehensive speech representation
\begin{equation}
\tilde{\mathbf{c}}_t = \left [ \mathbf{c}_t^T, \mathbf{r}_t^T \right ]^T
\end{equation}
for each time step. 
The sequence of the concatenated representations $\tilde{\mathbf{c}}_1, \ldots, \tilde{\mathbf{c}}_T$ is lastly fed into a prediction network (e.g., a linear projection) for the ultimate recognition task.

\subsection{Training Relational Thinking based Models}
%

For sequence modeling tasks like speech recognition, a common challenge arises from the varying lengths of the input and output sequences. This requires a loss function capable of managing such variations in sequence lengths. While \cite{huang2020deep} and \cite{chung2015recurrent} only considered the scenarios where the input and output sequences have equal lengths, 
our proposed spectro-temporal relational thinking framework is designed to handle scenarios where the input and output sequences can have varying lengths. 
However, a tractable loss function is required to enable the training of our proposed framework. Given the complexity introduced by the random processes governing the generation of the graph edges (\ref{eq.bern})--(\ref{eq.graph_embedding}), direct optimization of the model log-likelihood $\log p(y | \mathcal{C})$ is infeasible. Instead, we employ the variational lower bound $\mathcal{L}$ \cite{sohn2015learning}, by optimizing which log-likelihood can be also maximized:
\begin{equation}\label{eq.lower_bound}
\begin{aligned}
\log p(y | \mathcal{C}) \geq & \mathbb{E}_{ q \left . \left ( \tilde{\mathcal{A}}, \mathcal{S} \right | \mathcal{C} \right ) } \left [ \log p \left ( y \left | \mathcal{C}, \tilde{\mathcal{A}}, \mathcal{S} \right ) \right . \right ] - \\
& \text{div} \left . \left ( q \left . \left ( \tilde{\mathcal{A}}, \mathcal{S} \right | \mathcal{C} \right ) \right \| p \left . \left ( \tilde{\mathcal{A}}, \mathcal{S} \right | \mathcal{C} \right ) \right ) = \mathcal{L},
\end{aligned}
\end{equation}
where $\text{div}(\cdot \| \cdot)$ represents the KL divergence. In our proposed framework, we have two sets of variational latent variables that require optimization: $\tilde{\mathcal{A}} = \{ \tilde{\mathbf{A}}_1, \ldots, \tilde{\mathbf{A}}_T \}$ and $\mathcal{S} = \{ \mathbf{S}_1, \ldots, \mathbf{S}_T \}$, representing the adjacency matrices of the summary graphs and the graph transformation variable matrices for all time steps, respectively. $q ( \tilde{\mathcal{A}}, \mathcal{S} | \mathcal{C} )$ denotes the approximate posterior for $p( \tilde{\mathcal{A}}, \mathcal{S}  | \mathcal{C}, y )$, while $p ( \tilde{\mathcal{A}}, \mathcal{S} | \mathcal{C} )$ represents the prior \cite{nan2023variational}. As can be observed, $\mathcal{L}$ consists of a prediction objective (first term) and a regularization objective (second term). The prediction objective guides the model to recover the target sequence $y$, while the regularization objective encourages the model to keep its posterior distribution close to the prior. 
For the case where input and output sequences have equal lengths \cite{huang2020deep, chung2015recurrent}, the prediction objective can be decomposed into a frame-wise form as $\mathbb{E}_{ q ( \tilde{\mathcal{A}}, \mathcal{S} | \mathcal{C} ) } [\log p ( y | \mathcal{C}, \tilde{\mathcal{A}}, \mathcal{S} ) ] = \sum_{t = 1}^T \mathbb{E}_{ q ( \tilde{\mathcal{A}}, \mathcal{S} | \mathcal{C} ) } [\log p ( y_t | \mathbf{C}_t, \tilde{\mathcal{A}}, \mathcal{S} ) ]$. However, it does not generalize to our case where input and output sequences are of different lengths. This forces us to recover $y$ using $\mathcal{C}, \tilde{\mathcal{A}}, \mathcal{S}$ throughout all time steps by optimizing $\sum_{B \in {F^{-1}(y)}} \prod_{t = 1}^T p ( b_t | \mathcal{C}, \tilde{\mathcal{A}}, \mathcal{S} )$, where $B = [b_1, \ldots, b_{T}]$ denotes an alignment between $\mathcal{C}$ and $y$, and $F$ is a mapping function that maps $B$ with the same length as $\mathcal{C}$ to the target sequence $y$ \cite{graves2006connectionist} (see (\ref{eq.loss_rt}) in Appendix~\ref{app.learning}). 

On the other hand, according to \cite{nan2023variational}, since $p ( \tilde{\mathcal{A}}, \mathcal{S} | \mathcal{C} ) = \prod_{t = 1}^T p ( \tilde{\mathbf{A}}_t, \mathbf{S}_t | \mathbf{C}_t )$, the regularization objective can be decomposed as $\sum_{t = 1}^T \text{div}( q ( \tilde{\mathbf{A}}_t, \mathbf{S}_t | \mathbf{C}_t ) \| p ( \tilde{\mathbf{A}}_t, \mathbf{S}_t | \mathbf{C}_t ) )$, where $q ( \tilde{\mathbf{A}}_t, \mathbf{S}_t | \mathbf{C}_t )$ and $p ( \tilde{\mathbf{A}}_t, \mathbf{S}_t | \mathbf{C}_t )$ denote the approximate posterior and prior for time step $t$, respectively. Given that each element $s_{i, j}^{(t)}$ of $\mathbf{S}_t$ is conditioned on the Binomial variable $\tilde{\alpha}_{i, j}^{(t)}$ for the same edge of the $t$-th summary graph $\tilde{\mathcal{G}}_t$ (as indicated by (\ref{eq.s})), we can further derive the KL divergence term for each time step $t$ as
\begin{align}
\nonumber & \text{div} \left ( \left . q \left ( \left . \tilde{\mathbf{A}}_t, \mathbf{S}_t \right | \mathbf{C}_t \right ) \right \| p \left ( \left . \tilde{\mathbf{A}}_t, \mathbf{S}_t \right | \mathbf{C}_t \right ) \right ) \\
\label{eq.kl} = & \sum_{(i, j) \in \tilde{\mathcal{E}}_t} \left \{ \text{div} \left ( \left . q \left ( \left. \tilde{\alpha}_{i, j}^{(t)} \right | \mathbf{C}_t \right ) \right \| \left . p \left ( \tilde{\alpha}_{i, j}^{(t)} \right | \mathbf{C}_t \right ) \right ) + \right . \\
\nonumber & \mathbb{E}_{ \left . q \left ( \tilde{\alpha}_{i, j}^{(t)} \right | \mathbf{C}_t \right ) } \left [ \text{div} \left ( \left . q \left ( \left . s_{i, j}^{(t)} \right | \tilde{\alpha}_{i, j}^{(t)}, \mathbf{C}_t \right ) \right \| \left . \left . p \left ( s_{i, j}^{(t)} \right | \tilde{\alpha}_{i, j}^{(t)}, \mathbf{C}_t \right ) \right ) \right ] \right \},
\end{align}
where $\tilde{\mathcal{E}}_t$ denotes the edge set of the $t$-th summary graph. 
Following the Theorem~2 in \cite{huang2020deep}, we can further derive the closed-form of KL divergences between two Binomial distributions in (\ref{eq.kl}) as
\begin{align}
\label{eq.kl_binomial} & \text{div} \left ( q \left ( \left. \tilde{\alpha}_{i, j}^{(t)} \right | \mathbf{C}_t \right ) \left \| p \left ( \left. \tilde{\alpha}_{i, j}^{(t)} \right | \mathbf{C}_t \right ) \right . \right ) \\
\nonumber < & m_{i, j}^{(t)} \log \frac{ m_{i, j}^{(t)} }{ m_{i, j}^{(t, 0)} } + \left ( 1 - m_{i, j}^{(t)} \right ) \log \frac{ 1 - m_{i, j}^{(t)} + \frac{ m_{i, j}^{(t) 2} }{2} }{ 1 - m_{i, j}^{(t, 0)} + \frac{ m_{i, j}^{(t, 0) 2} }{2} },
\end{align}
where $m_{i, j}^{(t)} = n^{(t)} \tilde{\lambda}_{i, j}^{(t)}$ and $m_{i, j}^{(t, 0)} = n^{(t)} \tilde{\lambda}_{i, j}^{(t, 0)}$.
Similarly, the KL divergences between two Gaussian distributions in (\ref{eq.kl}) can be readily simplified to the following closed-form:
\begin{align}
\nonumber & \text{div} \left ( q \left ( \left . s_{i, j}^{(t)} \right | \tilde{\alpha}_{i, j}^{(t)}, \mathbf{C}_t \right ) \left \| p \left ( \left . s_{i, j}^{(t)} \right | \tilde{\alpha}_{i, j}^{(t)}, \mathbf{C}_t \right ) \right . \right ) \\
\label{eq.kl_gaussian} = & \frac{1}{2} \log \frac{ \sigma_{i, j}^{(t, 0) 2} }{ \sigma_{i, j}^{(t) 2} } + \frac{ \sigma_{i, j}^{(t) 2} + \left ( \mu_{i, j}^{(t)} - \mu_{i, j}^{(t, 0)} \right )^2 }{ 2 \sigma_{i, j}^{(t, 0) 2} } - \frac{1}{2}.
\end{align}
Finally, by substituting (\ref{eq.kl})--(\ref{eq.kl_gaussian}) into (\ref{eq.lower_bound}), we obtain a computationally tractable form of loss function, allowing direct optimization of the variational lower bound $\mathcal{L}$ of the model log-likelihood (more details are provided in Appendix~\ref{app.learning}).

\section{Experimental Configurations}\label{sec.experiment_settings}
\subsection{Dataset}
We evaluate our proposed acoustic modeling framework in a general phoneme recognition downstream task. 
The TIMIT dataset \cite{garofolo1993darpa} is used for training and evaluation, since it provides precise annotations for the start and end instants of each phoneme within an utterance, allowing for comprehensive analyses that 
lead to an in-depth understanding of what relations the proposed models learn and how their learning differs across various phoneme groups, in order to understand the benefits of the proposed mechanism.
To recover the target phoneme sequence $y$, we use the best path decoding method \cite{graves2006connectionist}. The phoneme error rate (PER) is employed for system evaluation. 

\subsection{Objectives}
To gain insights into how the proposed model could aid downstream tasks, we aim to answer the following questions:

\begin{tcolorbox}
\noindent {\bf Q1:}
{\it Does the proposed joint spectro-temporal modeling provide additional information that further benefits downstream tasks when compared to pure temporal or spectral modeling?}
 
\noindent {\bf Q2:}
{\it Is it more beneficial to model a larger context in the time domain or frequency domain?}

\noindent {\bf Q3:}
{\it Does the temporal span for relational thinking modeling affect the model's performance in downstream tasks?}

\noindent {\bf Q4:}
{\it Does relational thinking provide additional benefits beyond what the attention mechanism has achieved for downstream tasks?}

\noindent {\bf Q5:}
{\it Does the proposed framework consistently offer advantages across different types of acoustic features?}

\noindent {\bf Q6:}
{\it What does relational thinking actually learn?}
\end{tcolorbox}

\subsection{Experimental Settings}\label{sec.settings}
To answer the above questions, we apply our proposed acoustic modeling framework (as illustrated by Fig.~\ref{fig.wav2vec2_with_relational_thinking}) to a general phoneme recognition downstream task as described below. 

\subsubsection{Acoustic Feature Extraction}
Since wav2vec2 is one of the state-of-the-art frameworks for extracting speech representations \cite{baevski2020wav2vec}, we employ the pre-trained wav2vec2 BASE\footnote{\href{https://huggingface.co/facebook/wav2vec2-base}{https://huggingface.co/facebook/wav2vec2-base}.} as the acoustic feature extraction module in our proposed models.

\subsubsection{Spectro-temporal Relational Thinking}\label{sec.setting_relational_thinking}
\begin{figure}[]
\centering
\includegraphics[width=0.48\textwidth]{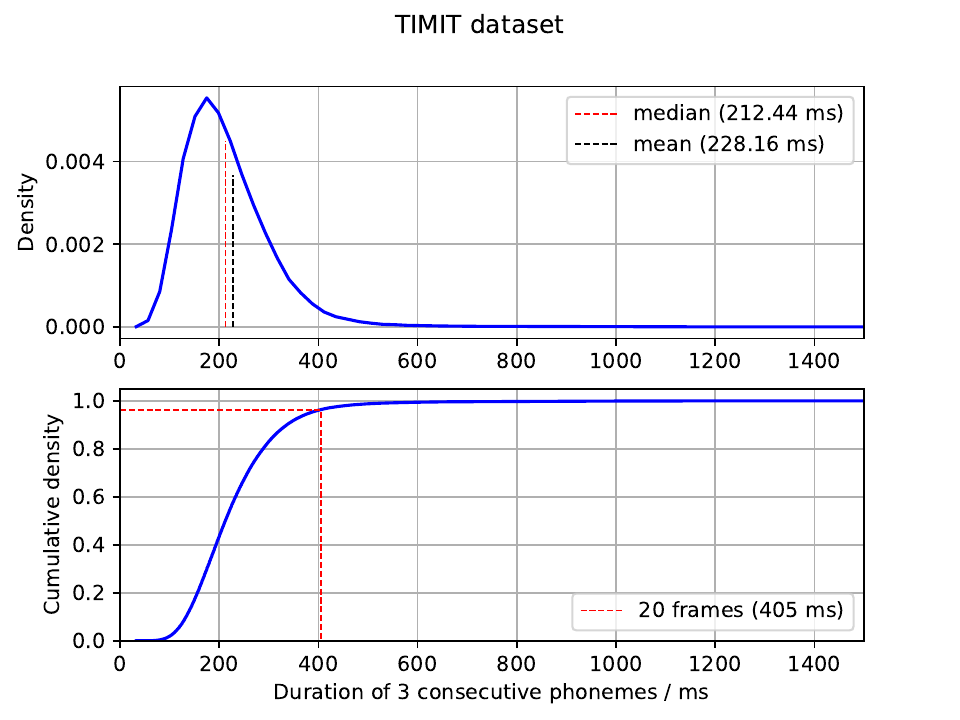}
\caption{Duration distribution of 3 consecutive phonemes within TIMIT dataset. Over 96\% of the tri-phone sequences have a duration shorter than 400 ms. 20 wav2vec2 frames cover a time span of 405 ms.}\label{fig.timit_duration_distribution}
\end{figure}

When modeling relational thinking for each time step $t$, it is essential to work with a feature map $\mathbf{C}_t$ that has a sufficiently wide context (i.e., with a sufficiently large $w$). This ensures that there is enough local context available for effective relational thinking modeling. Therefore, we take into account a context spanning at least 3 consecutive phonemes. This is in line with tri-phone models employed in HMM based acoustic models in the past \cite{jurafskyspeech}. 
We first investigate the duration distribution of 3 consecutive phonemes in the TIMIT dataset \cite{garofolo1993darpa}. As shown in Fig.~\ref{fig.timit_duration_distribution}, the majority (over 96\%) of these tri-phone sequences have a duration shorter than 400 ms. As the wav2vec2 framework uses a frame width of 25 ms and a frame stride of 20 ms, to create a feature map that is capable of effectively modeling relations among 3 consecutive phonemes for the majority of cases, it needs to consist of at least $w = 20$ frames, covering a time span of 405 ms. Hence, the feature map is designed as $\mathbf{C}_{t} = [\mathbf{c}_{t - 19}, \ldots, \mathbf{c}_t] \in \mathbb{R}^{768 \times 20}$, where 768 corresponds to the dimension of the context representations generated by wav2vec2 BASE. The kernel width and kernel stride for the temporal convolution in (\ref{eq.sub_sample}) are set to 5 and 2, respectively, leading to $\check{\mathbf{C}}_t \in \mathbb{R}^{768 \times 8}$ and the number of nodes included in the percept graphs being $u = 8$. Therefore, we can derive four different sets of resolution settings $(D^{(t)}, D^{(f)})$ for time and frequency domains, i.e., (8, 1), (4, 2), (2, 4), and (1, 8), respectively. 

For the approximate posterior $q ( \tilde{\alpha}_{i, j}^{(t)} | \mathbf{C}_t )$, we sample $\tilde{\alpha}_{i, j}^{(t)}$ from the Gaussian proxy $\mathcal{N} ( m_{i, j}^{(t)}, m_{i, j}^{(t)} (1 - m_{i, j}^{(t)} ) )$ of $\mathcal{B} ( n^{(t)}, \tilde{\lambda}_{i, j}^{(t)} )$ by letting $m_{i, j}^{(t)} = n^{(t)} \tilde{\lambda}_{i, j}^{(t)}$. However, as per Theorem~\ref{theorem_1}, before calculating the parameter $m_{i, j}^{(t)}$ of the Gaussian proxy with (\ref{eq.m}), we have to first learn the Gaussian distribution $\mathcal{N} ( \tilde{\mu}_{i, j}^{(t)}, \tilde{\sigma}_{i, j}^{(t) 2} )$ with $\tilde{\mu}_{i, j}^{(t)} < 1 / 2$  from the input $\mathbf{C}_t$. We thus employ two multi-layer perceptrons (MLPs) for the inference of $\tilde{\mu}_{i, j}^{(t)}$ and $\tilde{\sigma}_{i, j}^{(t)}$, respectively, taking $\mathbf{C}_t$ as their inputs. Each MLP has a hidden layer with 128 nodes. 
For the corresponding prior $p ( \tilde{\alpha}_{i, j}^{(t)} | \mathbf{C}_t )$, we learn the parameter $m_{i, j}^{(t, 0)}$ with an MLP (with 128 nodes in the hidden layer), taking as input the feature map $\mathbf{C}_{t}$. Note that we cannot directly draw samples $\tilde{\alpha}_{i, j}^{(t)}$ from the Gaussian proxy $\mathcal{N}( m_{i, j}^{(t)}, m_{i, j}^{(t)} (1 - m_{i, j}^{(t)} ) )$ here,  
and instead, re-parameterize \cite{kingma2013auto}. Specifically, we first draw an auxiliary variable $\gamma$ from $\mathcal{N}(0, 1)$. Then, we obtain $\tilde{\alpha}_{i, j}^{(t)}$ as $\tilde{\alpha}_{i, j}^{(t)} = m_{i, j}^{(t)} + (m_{i, j}^{(t)} ( 1 - m_{i, j}^{(t)} ) )^{1 / 2} \gamma$, enabling the parameter $m_{i, j}^{(t)}$ to be differentiable.
For the approximate posterior $q ( s_{i, j}^{(t)} | \tilde{\alpha}_{i, j}^{(t)}, \mathbf{C}_t )$, i.e., $\mathcal{N} ( \tilde{\alpha}_{i, j}^{(t)} \mu_{i, j}^{(t)}, \tilde{\alpha}_{i, j}^{(t)} \sigma_{i, j}^{(t)2} )$, we adopt two MLPs (with 128 nodes in the hidden layer) to predict $\mu_{i, j}^{(t)}$ and $\sigma_{i, j}^{(t)}$, respectively, taking $\mathbf{C}_t$ as inputs. The parameters of the corresponding prior, $\mu_{i, j}^{(t, 0)}$ and $\sigma_{i, j}^{(t, 0)}$, can be obtained similarly. Again, we rely on re-parameterization to sample $s_{i, j}^{(t)}$.

The node pair embedding function $\bar{f}_{\theta}(\cdot, \cdot)$ in (\ref{eq.graph_embedding}) is implemented with an MLP, where the hidden layer has 128 nodes. Context representations with respect to the two nodes are concatenated and then fed into the MLP. The output dimension of $\bar{f}_{\theta}(\cdot, \cdot)$ is 32. As a result, we eventually obtain a graph embedding $\mathbf{r}_t \in \mathbb{R}^{32}$ for each time step, together with the concatenated representation $\tilde{\mathbf{c}}_t \in \mathbb{R}^{800}$.

\subsubsection{Prediction Network}
Following \cite{baevski2020wav2vec}, a linear projection is added on top of $\tilde{\mathbf{c}}_1, \ldots, \tilde{\mathbf{c}}_T$ for the final recognition task. In line with the protocol outlined in \cite{lee1989speaker}, we keep all the 62 original phoneme classes during training, but collapse them to 39 classes during evaluation.

\section{Experimental Results and Analyses}
\subsection{Phoneme Recognition Performance}\label{sec.performance}
\subsubsection{Temporal vs. Spectral vs. Spectro-temporal}
We compare the performances of four proposed models, namely, t8f1, t2f4, t4f2, and t1f8, each adopting one of the four different resolution settings for time and frequency domains, respectively, as described in Section~\ref{sec.settings}. The models are named following the format ``$\text{t} D^{(t)} \text{f} D^{(f)}$''. Therefore, t2f4 and t4f2 correspond to the joint spectro-temporal modeling, while t8f1 and t1f8 in fact correspond to the temporal-only modeling and spectral-only modeling within a single domain, respectively.
We first use the pre-trained parameters within the wav2vec2 module to eliminate the impact of variations in acoustic features and solely evaluate the impact of joint spectro-temporal modeling. 
As shown in Table~\ref{tab.performance_wo_finetuning}, the two joint spectro-temporal models, t4f2 and t2f4, outperform the temporal and spectral models, t8f1 and t1f8. This comparison clearly demonstrates the advantage of joint spectro-temporal modeling over the temporal or spectral modeling within a single domain. It is also evident that all the proposed relational thinking models 
outperform the baseline model, the wav2vec2 BASE model
, with a relative reduction in PER ranging from 11.17\% to 19.61\%. 

\begin{table}[]
\centering
\caption{Phoneme recognition performances of baseline and proposed models without fine-tuning in terms of PER (\%).}\label{tab.performance_wo_finetuning}
\resizebox{0.48\textwidth}{!}{%
\begin{tabular}{cccccrr}
\toprule[1.5pt]
& & temporal & spectro & spectro-temporal & \multicolumn{1}{c}{\textbf{dev}} & \multicolumn{1}{c}{\textbf{test}} \\
\midrule[1pt]
baseline & wav2vec2 BASE & & & & 17.92 & 25.70 \\
\midrule[1pt]
\multirow{5}{*}{proposed} & t8f1 & \checkmark & & & 19.32 & 22.83 \\
 & t1f8 & & \checkmark & & 16.14 & 21.76 \\
 & t4f2 & & & \checkmark & 17.31 & \textbf{20.80} \\
 & t2f4 & & & \checkmark & 14.02 & \textbf{20.66} \\
 \cmidrule[1pt]{2-7}
 & w8-t2f4 & & & \checkmark & 18.89 & 22.93 \\
\bottomrule[1.5pt]
\end{tabular}%
}
\end{table}

\subsubsection{Trading off Temporal Context against Spectral Context}
We compare the models with a higher resolution in frequency domain to those with a higher resolution in time domain. Specifically, we compare t1f8 to t8f1, both of which model relations within a single (time or frequency) domain, and t2f4 to t4f2, which model relations in both time and frequency domains. As illustrated by Table~\ref{tab.performance_wo_finetuning}, in both comparisons, the model with a higher frequency domain resolution (t2f4 or t1f8) exhibit superiority over its counterpart with higher time domain resolution (t4f2 or t8f1). 
This suggests that there might be potential benefits in modeling relations across frequency bands in greater detail by setting a higher frequency domain resolution compared to focusing more on time domain relations.

\subsubsection{Impact of Temporal Span}
To further understand the impact of the temporal span for relational thinking modeling, i.e., the value of $w$ for every $\mathbf{C}_t$, on the performance of downstream task, we compare two proposed models with relational thinking modeled throughout 20 and 8 consecutive time steps, respectively. In other words, relational thinking is performed throughout temporal spans corresponding to tri-phones and mono-phones in the two models, respectively. We set the time and frequency resolutions to (2, 4) for both models. As shown in Table~\ref{tab.performance_wo_finetuning}, the w8-t2f4 model, which incorporates relational information only associated with the current phoneme at each time step, suffers a 10.99\% drop in performance when compared to the t2f4 model, which incorporates relational information associated with the current and 2 preceding phonemes. This suggests that certain spectro-temporal patterns associated with consecutive phonemes contribute to further improving the prediction performance for the current phoneme. However, w8-t2f4 still outperforms the wav2vec2 BASE baseline with a 10.78\% reduction in PER, validating the benefit of relational thinking modeling.

\begin{table}[]
\centering
\caption{Phoneme recognition performances of baselines and proposed models in terms of PER (\%) over TIMIT dataset.}\label{tab.performance}
\begin{tabular}{ccrr}
\toprule[1.5pt]
& & \multicolumn{1}{c}{\textbf{dev}} & \multicolumn{1}{c}{\textbf{test}} \\
\midrule[1pt]
\multirow{6}{*}{baseline} & CNN + TD-filterbanks \cite{zeghidour2018learning} & 15.60 & 18.00 \\
& PASE+ \cite{ravanelli2020multi} & -- & 17.20 \\
& Li-GRU + fMLLR \cite{ravanelli2018light} & -- & 14.90 \\
& wav2vec \cite{schneider2019wav2vec} & 12.90 & 14.70 \\
& vq-wav2vec \cite{baevski2019vq} & 9.60 & 11.60 \\
& wav2vec2 BASE \cite{baevski2020wav2vec} & 7.26 & 9.98 \\
\midrule[1pt]
\multirow{2}{*}{proposed} & t4f2 & 6.18 & \textbf{9.26} \\
& t2f4 & 6.23 & \textbf{9.20} \\
\bottomrule[1.5pt]
\end{tabular}
\end{table}

\subsubsection{Comparison with SOTA}
The proposed models are compared with the transformer (more essentially, self-attention mechanism) based wav2vec2 BASE baseline \cite{baevski2020wav2vec} and other state-of-the-art systems \cite{zeghidour2018learning, ravanelli2020multi, ravanelli2018light, schneider2019wav2vec, baevski2019vq} in Table~\ref{tab.performance}. To enable a fair comparison, we fine-tune the (wav2vec2) feature extraction module for both the proposed models and the baseline. 
Our proposed spectro-temporal models, t4f2 and t2f4, significantly outperform all the counterparts, specifically yielding 7.21\% and 7.82\% relative improvements in PER over the wav2vec2 BASE baseline in the test dataset, respectively, revealing the additional advantages offered by relational thinking modeling compared to self-attention mechanism in enhancing speech representation. 

\subsubsection{Generalization to Other Acoustic Features}
We also train a relational thinking based model using MFCCs (referred to as MFCC-RT-t2f4) and compare it with an MFCC baseline implemented with a simple linear projection (refer to Appendix~\ref{app.mfcc_models} for detailed configurations of the two models). As shown in Table~\ref{tab.mfcc}, the proposed MFCC-RT-t2f4 model significantly outperforms the MFCC baseline, achieving a 14.36\% reduction in PER over the test set. This validates that our proposed relational thinking modeling can generalize to sequential inputs composed of various types of acoustic features, providing additional relational information that consistently benefits downstream tasks.

\begin{table}[]
\centering
\caption{Phoneme recognition performances of baseline and proposed model using MFCCs.}\label{tab.mfcc}
\begin{tabular}{ccrr}
\toprule[1.5pt]
 &  & \multicolumn{1}{c}{\textbf{dev}} & \multicolumn{1}{c}{\textbf{test}} \\
\midrule[1pt]
baseline & MFCC & 39.80 & 47.90 \\
proposed & MFCC-RT-t2f4 & 39.58 & \textbf{41.02} \\
\bottomrule[1.5pt]
\end{tabular}
\end{table}

\subsection{Learned Relational Information}\label{sec.learned_relational_information}

In this subsection, we answer Q6 by further investigating what the proposed models actually learn in terms of the captured relational information and how it varies across phoneme sub groups (e.g., vowels, fricatives, etc.). We analyze model learning using a frame-wise phoneme classification task instead of the phoneme recognition task. This enables the analysis to focus solely on the relational information involved in every decomposed phoneme with no impact from neighboring phonemes in the sequence.
To be specific, given an arbitrary waveform (from the TIMIT dataset), for each segment aligned with a phoneme in the corresponding target sequence, we select 8 consecutive frames from its middle portion, and calculate the MFCCs $\mathbf{X}_{\text{MFCC}} = [ \mathbf{x}_{\text{MFCC}}^{(i - 7)}, \ldots, \mathbf{x}_{\text{MFCC}}^{(i)} ]$ as input for the phoneme classification task. The objective of this task is to predict the phoneme class of $\mathbf{X}_{\text{MFCC}}$. This process is feasible since the TIMIT dataset provides annotations for the start and end instants of every phoneme within an utterance. We use the proposed spectro-temporal relational thinking module to calculate the graph embedding $\mathbf{r}$, followed by an MLP for predicting the phoneme class using $\tilde{\mathbf{x}}_{\text{MFCC}}^{(i)} = [ \mathbf{x}_{\text{MFCC}}^{(i) T}, \mathbf{r}^T ]^T$. 

\begin{figure}[]
\centering
\includegraphics[width=0.48\textwidth]{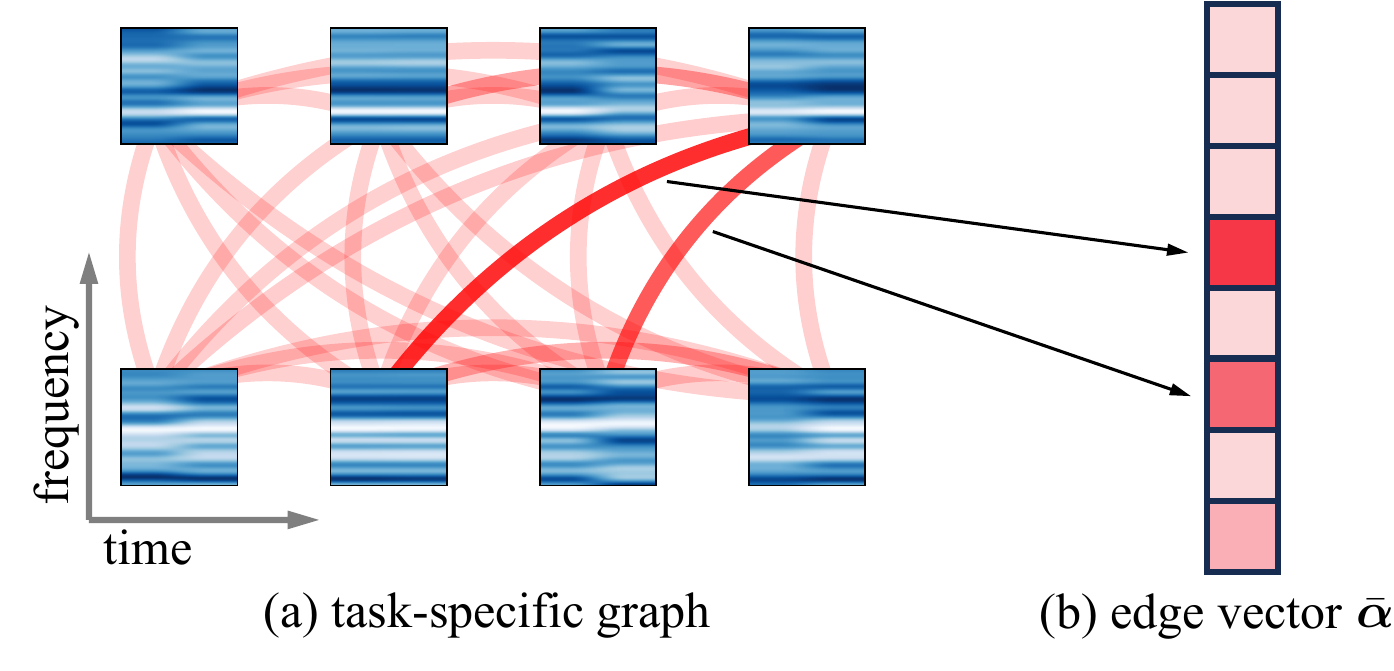}
\caption{Relations learned by spectro-temporal relational thinking. (a)~Relational thinking evaluates the importance of the co-occurrence of a pair of nodes, representing a novel type of information. A red curve represents an edge $\bar{\alpha}_{i, j}$ that connects a specific pair of sub feature maps. The intensity of an edge’s color corresponds to the regularized value of $\bar{\alpha}_{i, j}$ ranging from 0 to 1. A pair of nodes is of more importance when the edge connecting them attains a larger value of $\bar{\alpha}_{i, j}$. (b)~Edge vector $\boldsymbol{\alpha}$ collects all the edges $\bar{\alpha}_{i, j}$ in a task-specific graph.}\label{fig.relations}
\end{figure}

For each sample of the phoneme classification task, i.e., a feature map $\mathbf{X}_{\text{MFCC}}$, we can derive a task-specific graph using the trained classifier. As illustrated by Fig.~\ref{fig.relations}~(a), this graph clearly reveals the intricate relations among different sub feature maps of $\mathbf{X}_{\text{MFCC}}$. Obviously, different node pairs in the graph attain varying values of edge $\bar{\alpha}_{i, j}$, indicating that certain spectro-temporal patterns, i.e., the co-occurrence of certain sub feature maps are more important to the ultimate task than others, which are less meaningful.

\begin{figure}[t]
\centering
\includegraphics[width=0.48\textwidth]{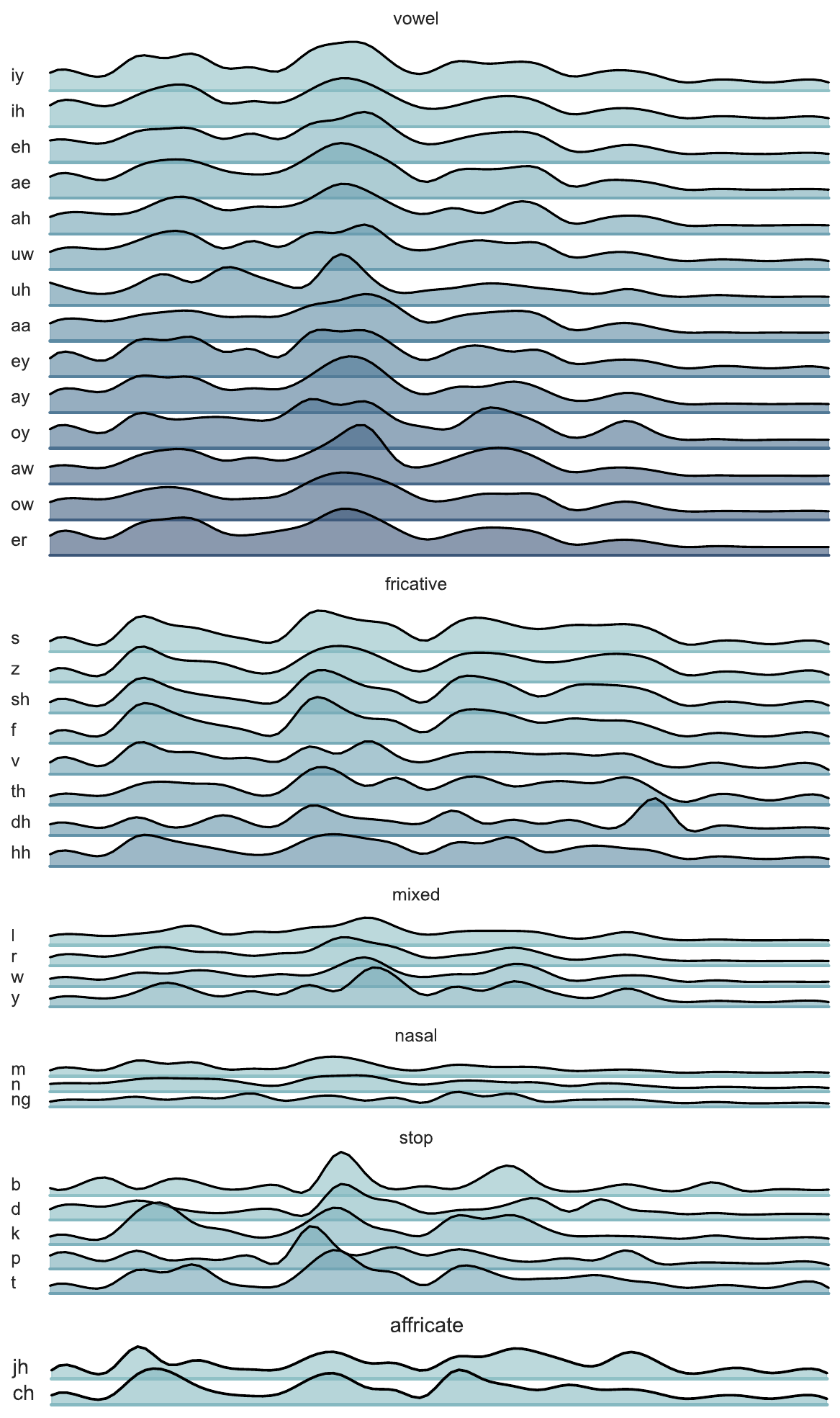}
\caption{Visualization of learned relational information. The mean of edge vectors $\bar{\boldsymbol{\alpha}}$ from each phoneme class is shown by groups. The captured relations exhibit more similarities for phoneme classes within the same sub group, but vary significantly between phoneme classes from different sub groups. The mixed group includes the approximants /w/ and /y/, as well as the liquids /l/ and /r/.}\label{fig.coefficient_phoneme_groups}
\end{figure}

Since the relational information is fully captured in the learned task-specific graphs, we next analyze these graphs for different phoneme sub groups. For the ease of comparison, we flatten the edges $\bar{\alpha}_{i, j}$ into a $\binom{8}{2} = 28$ dimensional edge vector $\bar{\boldsymbol{\alpha}}$ for each graph. This process is illustrated by Fig.~\ref{fig.relations}~(b). We visualize the mean of edge vectors from each phoneme class by groups in Fig.~\ref{fig.coefficient_phoneme_groups}. 
The captured relations 
i.e., the edges $\bar{\alpha}^{(t)}_{i, j}$
, exhibit more similarities for phoneme classes within the same sub group, but vary significantly between phoneme classes from different sub groups. This suggests that the proposed relational thinking modeling contributes to discerning and learning the intrinsic characteristics of various phoneme classes. 

\begin{figure}[]
\centering
\includegraphics[width=0.48\textwidth]{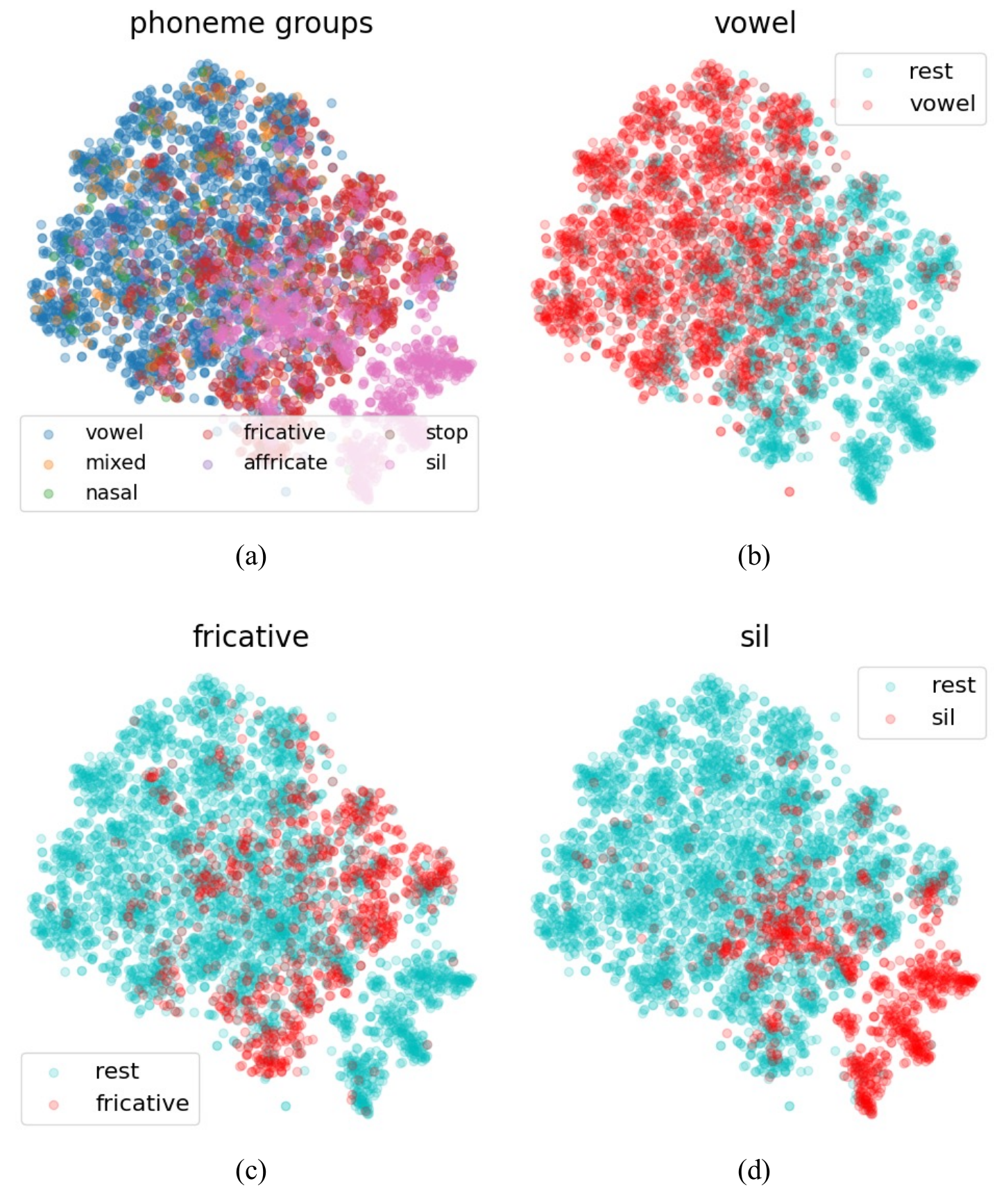}
\caption{t-SNE results for edge vectors. (a) t-SNE result for edge vectors from all phoneme groups. (b)--(d) t-SNE results for edge vectors from vowel/fricative/sil vs. edge vectors from the rest phoneme groups, respectively.}\label{fig.tsne_phoneme_groups}
\end{figure}

\subsubsection{Clustering of Edge Vectors}\label{sec.cluster_analysis}
We further cluster the edge vectors, aiming to understand whether the learned relational information correlates within each phoneme sub group and differentiates from other sub groups. We compute the t-SNE of the edge vectors $\bar{\boldsymbol{\alpha}}$ obtained from all the samples and visualize the clusters \cite{van2008visualizing}, as shown in Fig.~\ref{fig.tsne_phoneme_groups}~(a). For phoneme sub groups with a sufficient number of samples (vowel, fricative, silence, as indicated in Table~\ref{tab.phoneme_classification_performance}), the edge vectors are significantly clustered in the two-dimensional embedding space. Separate visualizations for each major phoneme sub group versus the rest are displayed in Fig.~\ref{fig.tsne_phoneme_groups} (b)--(d). Edge vectors from the minority sub groups with limited number of samples do not show prominent aggregations. Nevertheless, we can still conclude that the relational information involved in the learned graphs reveal similarities within phoneme sub groups and distinctions between phoneme sub groups.

\subsubsection{Classification Using Edge Vectors}
\begin{table}[]
\centering
\caption{Performance of phoneme group classification with edge vectors in terms of precision (\%).}\label{tab.phoneme_classification_performance}
\begin{tabular}{c r r}
\toprule[1.5pt]
\multicolumn{1}{c}{\textbf{phoneme group}} & \multicolumn{1}{c}{\textbf{precision}} & \multicolumn{1}{c}{\textbf{number of samples}} \\
\midrule[1pt]
vowel & \textbf{84.69} & \textbf{15001} \\
fricative & \textbf{81.37} & \textbf{6538} \\
sil & \textbf{93.36} & \textbf{6748} \\
affricate & 0 & 252 \\
mixed & 0.88 & 1582 \\
nasal & 0 & 700 \\
stop & 0 & 308 \\
\bottomrule[1.5pt]
\end{tabular}
\end{table}

Given the promising clustering results, we further conduct a classification analysis by training a simple MLP classifier using the edge vectors $\bar{\boldsymbol{\alpha}}$ as inputs. This classifier predicts the phoneme sub group for an input edge vector, with results shown in Table~\ref{tab.phoneme_classification_performance}. For the major phoneme sub groups with a sufficient number of samples, the classifier consistently achieves high precision. However, due to significant group imbalance, it struggles to correctly identify samples from the minority phoneme sub groups. Even though various specialized approaches exist to mitigate the group imbalance problem \cite{johnson2019survey}, addressing this issue was beyond the scope of our evaluation and left as future work. 
Our classification analysis further demonstrates that the learned edge vectors are indeed effective in distinguishing between phoneme sub groups, further validating the joint spectro-temporal modeling approach in capturing salient relations in speech.

\subsection{Analysis of Performance of Different Phoneme Groups}\label{sec.source_of_gain}

\begin{figure*}[]
\centering
\includegraphics[width=0.95\textwidth]{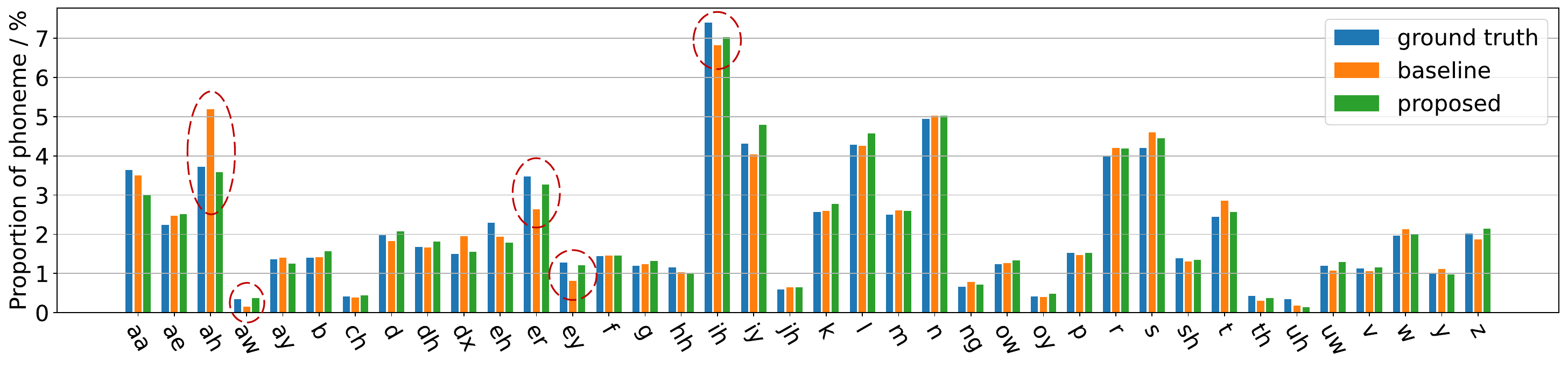}
\captionsetup{font=small}
\caption{Proportions of recognized phoneme classes by baseline and proposed w20-t2f4 model. Ground truth reveals the actual proportions of all phoneme classes in the TIMIT test set. The proportions of vowel classes recognized by the proposed model align more closely with the ground truth proportions, suggesting the proposed model's better performance in recognizing vowels. 
}\label{fig.proportion_of_phoneme}
\end{figure*}

In this subsection, we analyze the relational information learnt in phoneme recognition tasks, to gain a deeper understanding of how relational information is learnt from a local context of 3 consecutive phonemes rather than just a single phoneme (as discussed in Section~\ref{sec.learned_relational_information}) and how it enhances phoneme recognition performance. To do so, we initially compare the proportions of each phoneme class among all the phonemes (in the test set) recognized by the wav2vec2 BASE baseline and the proposed t2f4 model. These proportions are depicted in Fig.~\ref{fig.proportion_of_phoneme}, where the ground truth proportions of phoneme classes are also provided. Fig.~\ref{fig.proportion_of_phoneme} shows that the proportions of vowel classes (e.g., /ah/, /aw/, /er/, /ey/, /ih/ as circled out) recognized by the proposed model are more consistent with the ground truth proportions, with an average absolute difference of 0.23 pp, while the baseline shows a much higher average absolute difference of 0.35 pp (refer to Appendix~\ref{sec.proportions} for details).

\begin{figure}[]
\centering
\includegraphics[width=0.48\textwidth]{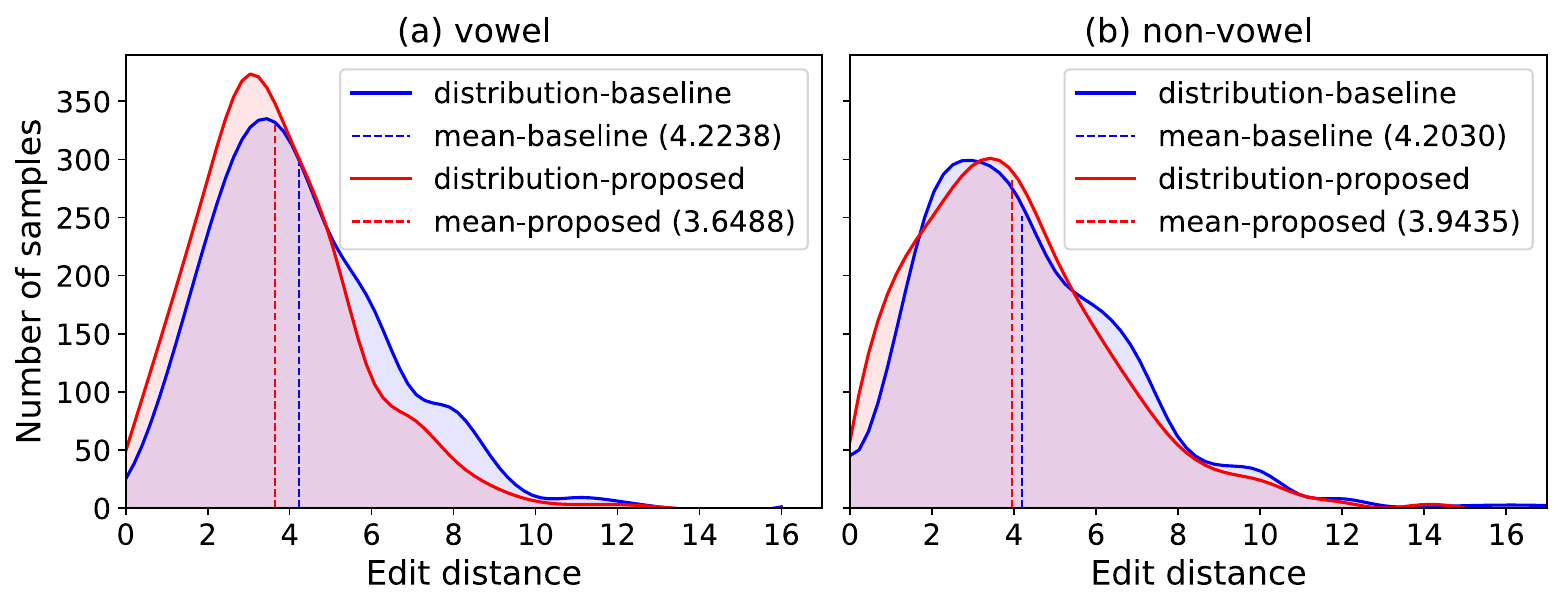}
\captionsetup{font=small}
\caption{Distributions of edit distances between recognized sequences and ground truth sequences. 
}\label{fig.edit_distance}
\end{figure}

We thus separately compare the errors made by both models in recognizing vowels and non-vowels.
To do this, given the recognition result of each model for a test sample, i.e., a sequence of recognized phonemes, we extract all the vowels/non-vowels from it and create a new sequence by combining the extracted phonemes with the original order preserved. This allows us to formulate recognized vowel/non-vowel sequences. For example, we can obtain a vowel sequence [/ix/, /ah/, /ix/, /ae/] from [/w/, /ix/, /dcl/, /s/, /ah/, /tcl/, /ch/, /ix/, /n/, /ae/]. The ground truth vowel/non-vowel sequences can be derived from the reference target sequence in the same way. To estimate the errors made by each model in recognizing vowels/non-vowels, we calculate the edit distance \cite{navarro2001guided} between the recognized vowel/non-vowel sequence and the corresponding ground truth counterpart for all test samples. 
Fig.~\ref{fig.edit_distance} illustrates the distributions of edit distances between the recognized sequences and the ground truth sequences for all test samples. In Fig.~\ref{fig.edit_distance}~(a), which pertains to the performances of the two models in recognizing vowels, it is evident that the proposed model outperforms the baseline. The distribution of edit distances for the proposed model is significantly skewed towards the left, compared to that for the baseline, with the average edit distance for the proposed model (3.6488) much smaller than that for the baseline (4.2238). While for the performances of the two models in recognizing non-vowels as depicted in Fig.~\ref{fig.edit_distance}~(b), the proposed model only shows a slight improvement over the baseline
, where the average edit distances for the two models are 3.9435 and 4.2030, respectively. This is potentially because vowel phonemes tend to have a longer duration than non-vowel phonemes, allowing the relational thinking module to capture more significant relational information within a local context with a fixed time span (405 ms), which in turn benefits the downstream task. Thus, we conclude that incorporating the biologically inspired relational thinking process benefits vowel recognition. This finding also aligns with the results of a speech intelligibility test conducted with human listeners \cite{meyer2006human}, suggesting that vowel identification is a relatively easier task for humans compared to consonant identification.

\begin{figure}[]
\centering
\includegraphics[width=0.48\textwidth]{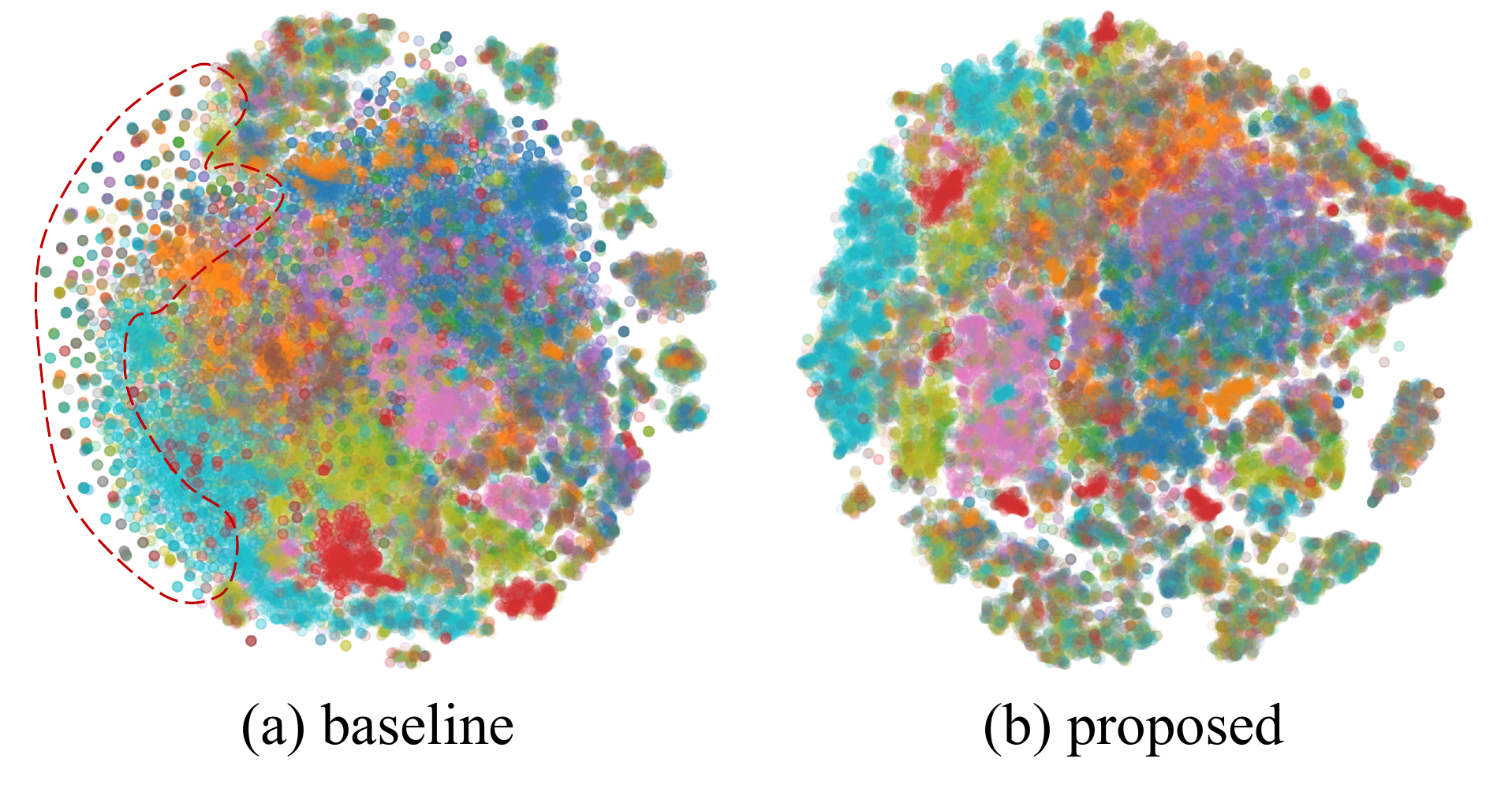}
\caption{t-SNE results for vowel latent vectors obtained from (a) baseline and (b) proposed model, respectively.}\label{fig.tsne}
\end{figure}

Lastly, we delve deeper into the analysis of the learned representation of vowels for both the baseline and the proposed model. The feature embeddings are extracted from the last layer of both models by selecting frames within the temporal span of any vowels in the test set. t-SNE is then applied to reduce the embeddings to two dimensions. The clusterings for the wav2vec2 baseline and the proposed t2f4 model are shown in Fig.~\ref{fig.tsne} (a) and Fig.~\ref{fig.tsne} (b), respectively.
There is a non-negligible proportion of scattered points (as circled out in Fig.~\ref{fig.tsne}~(a)), which are distant from any clusters and are interleaved with each other in a sparse region of the 2-dimensional embedding space. This suggests that the baseline may struggle to correctly classify the vowel frames corresponding to these data points, which is likely because the representations abstracted by the last layer of the baseline lack sufficient information to distinguish between all vowel classes. On the contrary, the proposed model benefits from the incorporation of additional relational information from the local context, with interleaving among data points from different vowel classes in the embedding space greatly reduced compared to that for the baseline. At the same time, data points from the same vowel class are still tightly clustered together, indicating better separability of vowel classes in the proposed model's representations.

\subsection{Speech Recognition with Proposed Framework}
\begin{table}[]
\centering
\caption{Speech recognition performances of baseline and proposed model in terms of WER (\%) over TIMIT test set.}\label{tab.speech_recognition_performance_timit}
\begin{tabular}{ccrr}
\toprule[1.5pt]
 &  & \multicolumn{1}{c}{\textbf{w/o LM}} & \multicolumn{1}{c}{\textbf{4-gram LM}} \\
\midrule[1pt]
baseline & wav2vec2 BASE \cite{baevski2020wav2vec} & 19.21 & 14.23 \\
proposed & w44-t4f2 & \textbf{18.72} & \textbf{13.77} \\
\bottomrule[1.5pt]
\end{tabular}
\end{table}

The proposed spetro-temporal relational thinking modeling is further validated in speech recognition tasks and evaluated using word error rate (WER) to demonstrate the generalizability of our proposed framework to other tasks. A word-level relational thinking model is built upon our proposed framework. In this model, we let $\mathbf{C}_{t} \in \mathbb{R}^{768 \times 44}$, spanning an average of 3 consecutive words. The kernel width and kernel stride for the temporal convolution in (\ref{eq.sub_sample}) are set to 9 and 5, respectively. The resolutions for time and frequency domains are set to (4, 2). As shown in Table~\ref{tab.speech_recognition_performance_timit}, this proposed model displays a 2.55\% reduction in WER against the wav2vec2 baseline \cite{baevski2020wav2vec} when language modeling is not applied. The incorporation of a 4-gram language model increases this reduction in WER to 3.23\%. These improvements imply that comprehending and utilizing the spectro-temporal relations associated with words also advantages the downstream speech recognition tasks as certain words tend to frequently appear together, such as ``I am''.

\section{Conclusion}\label{sec.conclusion}
We propose a novel spectro-temporal relational thinking based acoustic modeling framework, where its core module is inspired by a fundamental human learning process. This framework is capable of capturing a unique form of pair-wise information, distinct from the assessment of individual nodes as performed by attention mechanism. Models constructed using this framework show state-of-the-art performance in phoneme recognition tasks. Further analysis conveys connections between the captured relations and phoneme groups, where the patterns involved in the relations exhibit more similarities for phoneme classes within the same group, while showing significant variations between phoneme classes from different groups. Our analysis also reveals that relational thinking modeling primarily enhances the model's ability to recognize vowels. 
Additionally, we demonstrate the generalizability of the proposed framework by applying other types of acoustic features and employing it for different downstream tasks, where relational thinking modeling consistently benefits downstream tasks. This study aims to pave a new pathway for integrating biologically inspired human learning processes into deep learning approaches, improving the model's capability in speech recognition and potentially its interpretability.

%

\begin{appendices}

\section{Training Relational Thinking based Models}\label{app.learning}
The variational CTC loss \cite{nan2023variational}

\begin{align}
\label{eq.loss_rt} \tilde{\mathcal{L}} = & \sum_{B \in {F^{-1}(y)}} \prod_{t = 1}^T p \left ( b_t \left | \mathcal{C}, \tilde{\mathcal{A}}, \mathcal{S} \right . \right ) \\
\nonumber & \quad \quad - \sum_{t = 1}^T \text{div} \left ( \left . q \left ( \left . \tilde{\mathbf{A}}_t, \mathbf{S}_t \right | \mathbf{C}_t \right ) \right \| p \left ( \left . \tilde{\mathbf{A}}_t, \mathbf{S}_t \right | \mathbf{C}_t \right ) \right )
\end{align}
is employed to optimize $\mathcal{L}$ in \eqref{eq.lower_bound}, where $B = [b_1, \ldots, b_{T}]$ denotes an alignment between $\mathcal{C}$ and $y$ \cite{graves2006connectionist}, $b_t \in \mathcal{W} \cup \{ - \}$, $\mathcal{W}$ is the target vocabulary, and $F$ maps the paths $B$ with the same length as $\mathcal{C}$ to the target sequence $y$ by first merging the consecutive duplicated labels into one and then discarding the blanks ``$-$''. According to \cite{nan2023variational}, since $p ( \tilde{\mathcal{A}}, \mathcal{S} | \mathcal{C} ) = \prod_{t = 1}^T p ( \tilde{\mathbf{A}}_t, \mathbf{S}_t | \mathbf{C}_t )$, the KL divergence term in (\ref{eq.lower_bound}) is decomposed into a frame-wise form as in (\ref{eq.loss_rt}), where $q ( \tilde{\mathbf{A}}_t, \mathbf{S}_t | \mathbf{C}_t )$ and $p ( \tilde{\mathbf{A}}_t, \mathbf{S}_t | \mathbf{C}_t )$ denote the approximate posterior and prior for time step $t$, respectively. 

As each element $s_{i, j}^{(t)}$ of $\mathbf{S}_t$ is conditioned on the Binomial variable $\tilde{\alpha}_{i, j}^{(t)}$ for the same edge of the $t$-th summary graph $\tilde{\mathcal{G}}_t$ (as indicated by (\ref{eq.s})), the KL divergence terms in (\ref{eq.loss_rt}) can be further derived as
\begin{align}
\label{eq.kl_full} & \text{div} \left ( \left . q \left ( \left . \tilde{\mathbf{A}}_t, \mathbf{S}_t \right | \mathbf{C}_t \right ) \right \| p \left ( \left . \tilde{\mathbf{A}}_t, \mathbf{S}_t \right | \mathbf{C}_t \right ) \right ) \\
\nonumber = & \sum_{(i, j) \in \tilde{\mathcal{E}}_t} \text{div} \left ( \left . q \left ( \left . \tilde{\alpha}_{i, j}^{(t)}, s_{i, j}^{(t)} \right | \mathbf{C}_t \right ) \right \| p \left ( \left . \tilde{\alpha}_{i, j}^{(t)}, s_{i, j}^{(t)} \right | \mathbf{C}_t \right ) \right ) \\
\nonumber = & \sum_{(i, j) \in \tilde{\mathcal{E}}_t} \text{div} \left ( \left . q \left ( \tilde{\alpha}_{i, j}^{(t)} \right | \mathbf{C}_t \right ) \left . q \left ( s_{i, j}^{(t)} \right | \tilde{\alpha}_{i, j}^{(t)}, \mathbf{C}_t \right ) \right \| \\
\nonumber & ~~~~~~~~~~~~~~ \left . \left . p \left ( \tilde{\alpha}_{i, j}^{(t)} \right | \mathbf{C}_t \right ) \left . p \left ( s_{i, j}^{(t)} \right | \tilde{\alpha}_{i, j}^{(t)}, \mathbf{C}_t \right ) \right ) \\
\nonumber = & \sum_{(i, j) \in \tilde{\mathcal{E}}_t} \int_{\tilde{\alpha}_{i, j}^{(t)}} \int_{s_{i, j}^{(t)}} \left . q \left ( \tilde{\alpha}_{i, j}^{(t)} \right | \mathbf{C}_t \right ) \left . q \left ( s_{i, j}^{(t)} \right | \tilde{\alpha}_{i, j}^{(t)}, \mathbf{C}_t \right ) \cdot \\
\nonumber & ~~~~~~~~~ \log \frac{ \left . q \left ( \tilde{\alpha}_{i, j}^{(t)} \right | \mathbf{C}_t \right ) \left . q \left ( s_{i, j}^{(t)} \right | \tilde{\alpha}_{i, j}^{(t)}, \mathbf{C}_t \right ) }{ \left . p \left ( \tilde{\alpha}_{i, j}^{(t)} \right | \mathbf{C}_t \right ) \left . p \left ( s_{i, j}^{(t)} \right | \tilde{\alpha}_{i, j}^{(t)}, \mathbf{C}_t \right ) } d \tilde{\alpha}_{i, j}^{(t)} d s_{i, j}^{(t)} \\
\nonumber = & \sum_{(i, j) \in \tilde{\mathcal{E}}_t} \left \{ \int_{\tilde{\alpha}_{i, j}^{(t)}} \int_{s_{i, j}^{(t)}} \left . q \left ( s_{i, j}^{(t)} \right | \tilde{\alpha}_{i, j}^{(t)}, \mathbf{C}_t \right ) d s_{i, j}^{(t)} \cdot \right . \\
\nonumber & \left . q \left ( \tilde{\alpha}_{i, j}^{(t)} \right | \mathbf{C}_t \right ) \log \frac{ \left . q \left ( \tilde{\alpha}_{i, j}^{(t)} \right | \mathbf{C}_t \right ) }{ \left . p \left ( \tilde{\alpha}_{i, j}^{(t)} \right | \mathbf{C}_t \right ) } d \tilde{\alpha}_{i, j}^{(t)} + \int_{\tilde{\alpha}_{i, j}^{(t)}} \left . q \left ( \tilde{\alpha}_{i, j}^{(t)} \right | \mathbf{C}_t \right ) \cdot \\
\nonumber & \left . \int_{s_{i, j}^{(t)}} \left . q \left ( s_{i, j}^{(t)} \right | \tilde{\alpha}_{i, j}^{(t)}, \mathbf{C}_t \right ) \log \frac{ \left . q \left ( s_{i, j}^{(t)} \right | \tilde{\alpha}_{i, j}^{(t)}, \mathbf{C}_t \right ) }{ \left . p \left ( s_{i, j}^{(t)} \right | \tilde{\alpha}_{i, j}^{(t)}, \mathbf{C}_t \right ) } d s_{i, j}^{(t)} d \tilde{\alpha}_{i, j}^{(t)} \right \} \\
\nonumber = & \sum_{(i, j) \in \tilde{\mathcal{E}}_t} \left \{ \text{div} \left ( \left . q \left ( \left. \tilde{\alpha}_{i, j}^{(t)} \right | \mathbf{C}_t \right ) \right \| \left . p \left ( \tilde{\alpha}_{i, j}^{(t)} \right | \mathbf{C}_t \right ) \right ) + \right . \\
\nonumber & \mathbb{E}_{ \left . q \left ( \tilde{\alpha}_{i, j}^{(t)} \right | \mathbf{C}_t \right ) } \left [ \text{div} \left ( \left . q \left ( \left . s_{i, j}^{(t)} \right | \tilde{\alpha}_{i, j}^{(t)}, \mathbf{C}_t \right ) \right \| \left . \left . p \left ( s_{i, j}^{(t)} \right | \tilde{\alpha}_{i, j}^{(t)}, \mathbf{C}_t \right ) \right ) \right ] \right \},
\end{align}
where
\begin{equation}
\begin{aligned}
q \left ( \left. \tilde{\alpha}_{i, j}^{(t)} \right | \mathbf{C}_t \right ) & = \mathcal{B} \left ( n^{(t)}, \tilde{\lambda}_{i, j}^{(t)} \right ), \\
p \left ( \left. \tilde{\alpha}_{i, j}^{(t)} \right | \mathbf{C}_t \right ) & = \mathcal{B} \left ( n^{(t)}, \tilde{\lambda}^{(t, 0)}_{i, j} \right ), \\
q \left ( \left . s_{i, j}^{(t)} \right | \tilde{\alpha}_{i, j}^{(t)}, \mathbf{C}_t \right ) & = \mathcal{N} \left ( \tilde{\alpha}_{i, j}^{(t)} \mu_{i, j}^{(t)}, \tilde{\alpha}_{i, j}^{(t)} \sigma_{i, j}^{(t)2} \right ), \\
p \left ( \left . s_{i, j}^{(t)} \right | \tilde{\alpha}_{i, j}^{(t)}, \mathbf{C}_t \right ) & = \mathcal{N} \left ( \tilde{\alpha}_{i, j}^{(t)} \mu_{i, j}^{(t, 0)}, \tilde{\alpha}_{i, j}^{(t)} \sigma_{i, j}^{(t, 0) 2} \right ).
\end{aligned}
\end{equation}

\section{Detailed Experimental Configurations for Models Using MFCCs}\label{app.mfcc_models}
To demonstrate the generalizability of our proposed framework, we also train a set of models using the MFCCs as acoustic features. The MFCC baseline is implemented with a simple linear projection, taking as input the MFCC feature vectors from all time steps, instead of the context representations generated by wav2vec2 BASE. In contrast, the relational thinking based MFCC model, referred to as MFCC-RT-t2f4, further computes the graph embeddings for all time steps using the feature maps $\mathbf{C}_t$ obtained from the MFCC feature vectors, following the procedures outlined in Section~\ref{seq.wav2vec2_with_rt}. Here, the width of $\mathbf{C}_t$ is set to $w = 20$, and the resolutions for time and frequency domains are (2, 4). The MFCC feature vectors and the graph embeddings are concatenated before fed into the linear projection. Since MFCC calculations are deterministic, fine-tuning is not required for these two models.

\section{Proportions of Phoneme Classes}\label{sec.proportions}
\begin{table}[]
\centering
\caption{Proportions (\%) of phoneme classes recognized by baseline and proposed models. Values within parentheses indicate the absolute differences (pp) between the proportions recognized by the two models and the ground truth proportions.}\label{tab.proportions}
\resizebox{0.48\textwidth}{!}{%
\begin{tabular}{ccrrr}
\toprule[1.5pt]
\multicolumn{2}{c}{\textbf{phoneme}} & \multicolumn{1}{c}{\begin{tabular}[c]{@{}c@{}}\textbf{recognized by}\\ \textbf{baseline}\end{tabular}} & \multicolumn{1}{c}{\begin{tabular}[c]{@{}c@{}}\textbf{recognized by}\\ \textbf{proposed}\end{tabular}} & \multicolumn{1}{c}{\textbf{ground truth}} \\
\midrule[1pt]
\multirow{15}{*}{vowel} & aa         & 3.50 (0.14)                                                                          & 3.01 (0.63)                                                                          & 3.64                             \\
& ae         & 2.47 (0.23)                                                                          & 2.51 (0.28)                                                                          & 2.24                             \\
& ah         & 5.18 (1.46)                                                                          & 3.59 (0.14)                                                                          & 3.72                             \\
& aw         & 0.15 (0.19)                                                                          & 0.37 (0.02)                                                                          & 0.34                             \\
& ay         & 1.40 (0.04)                                                                          & 1.26 (0.10)                                                                          & 1.35                             \\
& eh         & 1.94 (0.35)                                                                          & 1.79 (0.50)                                                                          & 2.29                             \\
& er         & 2.63 (0.84)                                                                          & 3.27 (0.20)                                                                          & 3.47                             \\
& ey         & 0.82 (0.46)                                                                          & 1.22 (0.07)                                                                          & 1.28                             \\
& ih         & 6.82 (0.58)                                                                          & 7.03 (0.37)                                                                          & 7.40                              \\
& iy         & 4.03 (0.27)                                                                          & 4.79 (0.48)                                                                          & 4.31                             \\
& ow         & 1.26 (0.02)                                                                          & 1.33 (0.10)                                                                          & 1.24                             \\
& oy         & 0.40 (0.02)                                                                          & 0.48 (0.06)                                                                          & 0.42                             \\
& uh         & 0.19 (0.17)                                                                          & 0.14 (0.21)                                                                          & 0.35                             \\
& uw         & 1.07 (0.12)                                                                          & 1.29 (0.09)                                                                          & 1.19                             \\
& \underline{average} & \underline{(0.35)} & \underline{(\textbf{0.23})} & \\
\midrule[1pt]
\multirow{26}{*}{non-vowel} & b          & 1.42 (0.01)                                                                          & 1.57 (0.16)                                                                          & 1.41                             \\
& ch         & 0.39 (0.02)                                                                          & 0.44 (0.03)                                                                          & 0.41                             \\
& d          & 1.83 (0.15)                                                                          & 2.07 (0.09)                                                                          & 1.98                             \\
& dh         & 1.66 (0.01)                                                                          & 1.82 (0.14)                                                                          & 1.67                             \\
& dx         & 1.96 (0.46)                                                                          & 1.55 (0.05)                                                                          & 1.49                             \\
& f          & 1.45 (0.00)                                                                          & 1.46 (0.01)                                                                          & 1.45                             \\
& g          & 1.23 (0.03)                                                                          & 1.32 (0.12)                                                                          & 1.20                              \\
& hh         & 1.03 (0.13)                                                                          & 1.01 (0.15)                                                                          & 1.15                             \\
& jh         & 0.64 (0.05)                                                                          & 0.64 (0.05)                                                                          & 0.59                             \\
& k          & 2.60 (0.04)                                                                          & 2.77 (0.21)                                                                          & 2.57                             \\
& l          & 4.26 (0.03)                                                                          & 4.58 (0.29)                                                                          & 4.29                             \\
& m          & 2.60 (0.10)                                                                          & 2.59 (0.09)                                                                          & 2.50                              \\
& n          & 5.02 (0.07)                                                                          & 5.03 (0.08)                                                                          & 4.95                             \\
& ng         & 0.79 (0.12)                                                                          & 0.71 (0.05)                                                                          & 0.67                             \\
& p          & 1.47 (0.05)                                                                          & 1.52 (0.00)                                                                          & 1.52                             \\
& r          & 4.21 (0.19)                                                                          & 4.19 (0.17)                                                                          & 4.01                             \\
& s          & 4.60 (0.40)                                                                          & 4.45 (0.25)                                                                          & 4.20                              \\
& sh         & 1.30 (0.08)                                                                          & 1.35 (0.03)                                                                          & 1.38                             \\
& sil        & 20.33 (0.02)                                                                         & 19.70 (0.61)                                                                         & 20.31                            \\
& t          & 2.86 (0.42)                                                                          & 2.57 (0.13)                                                                          & 2.44                             \\
& th         & 0.30 (0.12)                                                                          & 0.37 (0.05)                                                                          & 0.42                             \\
& v          & 1.05 (0.07)                                                                          & 1.15 (0.02)                                                                          & 1.13                             \\
& w          & 2.13 (0.16)                                                                          & 1.99 (0.02)                                                                          & 1.97                             \\
& y          & 1.11 (0.10)                                                                          & 0.97 (0.03)                                                                          & 1.01                             \\
& z          & 1.87 (0.15)                                                                          & 2.14 (0.12)                                                                          & 2.02                             \\
& \underline{average} & \underline{(0.12)} & \underline{(0.12)} & \\
\bottomrule[1.5pt]
\end{tabular}%
}
\end{table}

Table~\ref{tab.proportions} presents the proportions of phoneme classes recognized by the wav2vec2 baseline and the proposed t2f4 model, along with the ground truth proportions of phoneme classes in the test set. Values within parentheses indicate the absolute differences between the proportions recognized by the two models and the ground truth proportions. The average absolute difference between the proportions of recognized vowel classes by the baseline and the ground truth is 0.35 pp. In contrast, the proposed model shows a much smaller average absolute difference among vowel classes against the ground truth, which is only 0.23 pp. While for non-vowel classes, both models produce an average absolute difference of approximately 0.12 pp between the proportions of recognized phoneme classes and the ground truths.

\end{appendices}

\bibliographystyle{IEEEtran}
\bibliography{references}

\vfill

\end{document}